\RequirePackage{amsmath}
\documentclass[runningheads]{llncs}

\usepackage[utf8]{inputenc}
\usepackage[T1]{fontenc}

\usepackage{amssymb}

\usepackage{xcolor}
\usepackage[hidelinks]{hyperref}

\newcommand{\op}[1]{\operatorname{#1}}

\usepackage{todonotes}

\usepackage{listings}
\usepackage{courier}
\lstdefinelanguage{tessla}{
  morekeywords={def,in,out},
  morecomment=[l]{\#},
  morekeywords={[2]merge,last},
  morekeywords={[3]Events,Unit}
}

\usepackage{stmaryrd} 
\usepackage{utf8-math}   
\usepackage{graphicx}

\usepackage{paralist} 

\usepackage{amsthm}

\usepackage{tikz}
\usetikzlibrary{arrows.meta}
\tikzset{
  event/.style={
    draw,
    inner sep=.5pt,
    circle,
    minimum width=6pt
  },
  events/.style={
    font=\tiny,
    xscale=.5,
    yscale=.3
  },
  unit/.style={
    event,
    path picture={ 
      \draw[black]
        (path picture bounding box.south east) -- (path picture bounding box.north west)
        (path picture bounding box.south west) -- (path picture bounding box.north east);
    }
  }
}

\allowdisplaybreaks[1] 

\newcommand{\ignore}[1]{}

\let\op\operatorname
\newcommand{\testf}{\textrm{t\kern-0.14em f}}
\newcommand{\testrue}{\textrm{t\kern-0.16em t}}
\newcommand{\tesfalse}{\textrm{f\kern-0.1em f}}

\newcommand{\streamincl}{\bot}
\newcommand{\unitsym}{\boxempty}

\newcommand{\sem}[1]{\llbracket #1 \rrbracket}
\let\phi\varphi
\let\epsilon\varepsilon

\newcommand{\signaldx}{.15}
\newcommand{\signaldy}{.35}
\newcommand{\firstbar}[4]{
  \pgfmathsetmacro{\signallength}{#3-\signaldx}
  \filldraw[draw=black,fill=#1]
    #2 ++(0,\signaldy) -- ++(\signallength,0) -- ++(\signaldx,-\signaldy)
    --++(-\signaldx,-\signaldy) -- ++(-\signallength,0) -- cycle;
  \pgfmathsetmacro{\halflength}{.5*#3}
  \path #2 ++(\halflength,0) node {#4};
}
\newcommand{\midbar}[4]{
  \pgfmathsetmacro{\signallength}{#3-2*\signaldx}
  \filldraw[draw=black,fill=#1]
    #2 -- ++(\signaldx,\signaldy) --++(\signallength,0) --++(\signaldx,-\signaldy)
    --++(-\signaldx,-\signaldy) --++(-\signallength,0) -- cycle;
  \pgfmathsetmacro{\halflength}{.5*#3}
  \path #2 ++(\halflength,0) node {#4};
}
\newcommand{\lastbar}[4]{
  \pgfmathsetmacro{\signallength}{#3-\signaldx}
  \fill[#1]
    #2 -- ++(\signaldx,\signaldy) -- ++(\signallength,0) --++(0,-\signaldy)
    --++(0,-\signaldy) --++(-\signallength,0) -- cycle;
  \draw[black]
    #2 -- ++(\signaldx,\signaldy) -- ++(\signallength,0)
    #2 -- ++(\signaldx,-\signaldy) -- ++(\signallength,0);
  \pgfmathsetmacro{\halflength}{.5*#3}
  \path #2 ++(\halflength,0) node {#4};
}

\makeatletter
\newenvironment{customTheorem}[1]
  {\count@\c@theorem
   \global\c@theorem#1 %
    \global\advance\c@theorem\m@ne
   \theorem}
  {\endtheorem
   \global\c@theorem\count@}
\makeatother

\spnewtheorem*{proofsketch}{Proof Sketch}{\itshape}{\rmfamily}

\begin{document}

\title{%
  TeSSLa: Temporal Stream-based\\ Specification Language%
  \thanks{This work is supported in part by the European COST Action ARVI,
    the BMBF project ARAMiS II with funding~ID 01~IS~16025,
    and the European Horizon 2020 project COEMS under number 732016.}
}

\titlerunning{TeSSLa: Temporal Stream-based Specification Language}

\author{Lukas Convent \and
  Sebastian Hungerecker \and
  Martin Leucker \and\\
  Torben Scheffel \and
  Malte Schmitz \and
  Daniel Thoma}

\authorrunning{L. Convent, S. Hungerecker, M. Leucker, T. Scheffel, M. Schmitz, D. Thoma}

\institute{Institute for Software Engineering, University of Lübeck, Lübeck, Germany\\
\email{\{convent,hungerecker,leucker,scheffel,schmitz,thoma\}\\@isp.uni-luebeck.de}}

\maketitle

\begin{abstract}
Runtime verification is concerned with monitoring program traces.
In particular, stream runtime verification (SRV) takes the program trace as input streams and incrementally derives output streams.
SRV can check logical properties and compute temporal metrics and statistics from the trace.
We present TeSSLa, a temporal stream-based specification language for SRV.
TeSSLa supports timestamped events natively and is hence suitable for streams that are both sparse and fine-grained, which often occur in practice.
We prove results on TeSSLa's expressiveness and compare different TeSSLa fragments to (timed) automata, thereby inheriting various decidability results.
Finally, we present a monitor implementation and prove its correctness.
\end{abstract}


\section{Introduction}

\noindent The essence of software verification is to check whether a program meets its specification.
Runtime verification (RV) is an applied formal technique that has been established as a complement to traditional verification techniques such as model checking~\cite{leucker09brief,havelund05verify}.
Compared to static verification, RV considers only a single run of a system and checks whether it satisfies a property.
Thus, RV can be seen as a lightweight, but formal extension to testing and debugging.
RV can be applied offline to previously recorded traces or online to evaluate correctness properties at the runtime of the system under scrutiny.
Typically, a property to be checked is specified as a logical formula, e.g. in (past time) LTL, and then synthesized to a monitor which can evaluate a run~\cite{havelund_synthesizing_2002,bauer_runtime_2011}.
Stream runtime verification (SRV)~\cite{bozzelli14foundations}, as pioneered by the language LOLA~\cite{LOLA,LOLA2}, takes a different approach by incrementally relating a set of input streams to a set of output streams.
This allows not only the monitoring of correctness properties but also of quantitative measures.
In this paper we introduce the novel temporal stream-based specification language TeSSLa which is tailored for SRV of cyber-physical systems, where timing is a critical issue.
While traditional SRV approaches process event streams without considering timing information, TeSSLa supports timestamped events natively, which allows efficient processing of streams with sparse and fine-grained event sequences.
Preliminary versions of TeSSLa have already been studied with regard to their usability to monitor trace data generated by embedded tracing units of processors~\cite{conirasdate}; how to implement stream-based monitors on hardware has been studied in theory~\cite{oldtessla} and practice~\cite{conirasrv}. These versions share the basic idea of transforming timed event streams but they did not allow for recursive equations and comprised only a set of ad-hoc operators. In this paper we define a minimal language with support for recursive definitions that allows us to obtain strong guarantees for evaluation algorithms, expressiveness results and meaningful fragments. While the practical applicability of such a language has been demonstrated by the previous papers, these papers lack a concise and clear theoretical basis and investigation.
As an example for SRV, consider the following specification which checks whether a measured temperature stays within given boundaries. For every new event (measurement) on the temperature stream, new events on the derived streams \emph{low}, \emph{high} and \emph{unsafe} are computed:
\begin{center}
  \begin{minipage}{6cm}
    \setlength{\belowdisplayskip}{0pt}\setlength{\abovedisplayskip}{1ex}
    \begin{align*}
      \mathit{low} &:= \mathit{temperature} < 3\\
      \mathit{high} &:= \mathit{temperature} > 8\\
      \mathit{unsafe} &:= \mathit{low} \vee \mathit{high}
    \end{align*}
  \end{minipage}
  \begin{minipage}{5cm}

\begin{tikzpicture}[events]
  \node[event, fill=yellow!10] at (1,0) (t1) {6};
  \node[event, fill=yellow!10] at (2,0) (t2) {2};
  \node[event, fill=yellow!10] at (3,0) (t3) {1};
  \node[event, fill=yellow!10] at (4,0) (t4) {5};
  \node[event, fill=yellow!10] at (5,0) (t5) {9};

  \node[event, fill=red!10] at (1,-1) (l1) {\tesfalse};
  \node[event, fill=green!50!black!10] at (2,-1) (l2) {\testrue};
  \node[event, fill=green!50!black!10] at (3,-1) (l3) {\testrue};
  \node[event, fill=red!10] at (4,-1) (l4) {\tesfalse};
  \node[event, fill=red!10] at (5,-1) (l5) {\tesfalse};

  \node[event, fill=red!10] at (1,-2) (h1) {\tesfalse};
  \node[event, fill=red!10] at (2,-2) (h2) {\tesfalse};
  \node[event, fill=red!10] at (3,-2) (h3) {\tesfalse};
  \node[event, fill=red!10] at (4,-2) (h4) {\tesfalse};
  \node[event, fill=green!50!black!10] at (5,-2) (h5) {\testrue};

  \node[event, fill=red!10] at (1,-3) (u1) {\tesfalse};
  \node[event, fill=green!50!black!10] at (2,-3) (u2) {\testrue};
  \node[event, fill=green!50!black!10] at (3,-3) (u3) {\testrue};
  \node[event, fill=red!10] at (4,-3) (u4) {\tesfalse};
  \node[event, fill=green!50!black!10] at (5,-3) (u5) {\testrue};

  \path
    (0,0) node[left] {\textit{temperature}} edge[|-]
    (t1) (t1) edge (t2) (t2) edge (t3) (t3) edge (t4) (t4) edge (t5) (t5) edge[->] +(1,0);
  \path
    (0,-1) node[left] {\textit{low}} edge[|-]
    (l1) (l1) edge (l2) (l2) edge (l3) (l3) edge (l4) (l4) edge (l5) (l5) edge[->] +(1,0);
  \path
    (0,-2) node[left] {\textit{high}} edge[|-]
    (h1) (h1) edge (h2) (h2) edge (h3) (h3) edge (h4) (h4) edge (h5) (h5) edge[->] +(1,0);
  \path
    (0,-3) node[left] {\textit{unsafe}} edge[|-]
    (u1) (u1) edge (u2) (u2) edge (u3) (u3) edge (u4) (u4) edge (u5) (u5) edge[->] +(1,0);
\end{tikzpicture}

  \end{minipage}
\end{center}
SRV is a combination of complex event processing (CEP) and traditional RV approaches: Streams are transformed into streams and there is not only one final verdict but the output is a stream of the property being evaluated at every temperature change. Furthermore, the user gets more detailed information about why an error occurred by being able to distinguish between the two separate causes \emph{low} and \emph{high}.

In the rest of this section we introduce the main features of TeSSLa and contrast them with related specification languages. The next section presents the language and its semantics formally, in \autoref{sec:properties} we present several results regarding the expressiveness of TeSSLa and in \autoref{sec:transducers} we focus on comparing (fragments of) the language to variants of (timed) automata. Finally in \autoref{sec:implementation} we discuss different approaches to implement TeSSLa monitors and present our TeSSLa tool suite.

\paragraph{Asynchronous Streams}
In the previous example of traditional SRV, every stream has an event for every step of the system.
TeSSLa requires the events of all streams to be in a global order, but doesn't require all streams to have simultaneous events.
As a consequence, both sparse and high-frequency streams can be modeled. As cyber-physical systems often give rise to streams at unstable frequencies or continuous signals, this asynchronous setting is especially suitable.
Consider as an example a ring buffer where the number of write accesses should not exceed the number of read accesses too much:

\begin{center}
  \begin{minipage}{6.9cm}
    \setlength{\belowdisplayskip}{0pt}\setlength{\abovedisplayskip}{1ex}
    \begin{align*}
      \mathit{numReads} &:= \mathbf{count}(\mathit{read})\\
      \mathit{numWrites} &:= \mathbf{count}(\mathit{write})\\
      \mathit{safe} &:= \mathit{numWrites} - \mathit{numReads} \le 2
    \end{align*}
  \end{minipage}
  \begin{minipage}{5.1cm}

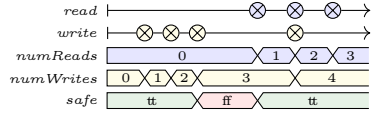
\begin{tikzpicture}[events]
  \node[unit, fill=blue!10] at (4,0) (r1) {};
  \node[unit, fill=blue!10] at (5,0) (r2) {};
  \node[unit, fill=blue!10] at (6,0) (r3) {};

  \path
    (0,0) node[left] {\textit{read}} edge[|-]
    (r1) (r1) edge (r2) (r2) edge (r3) (r3) edge[->] +(1,0);

  \node[unit, fill=yellow!10] at (1,-1) (w1) {};
  \node[unit, fill=yellow!10] at (1.7,-1) (w2) {};
  \node[unit, fill=yellow!10] at (2.4,-1) (w3) {};
  \node[unit, fill=yellow!10] at (5,-1) (w4) {};

  \path
    (0,-1) node[left] {\textit{write}} edge[|-]
    (w1) (w1) edge (w2) (w2) edge (w3) (w3) edge (w4) (w4) edge[->] +(2,0);
  
  \node[left] at (0,-2) {\textit{numReads}};
  \firstbar{blue!10}{(0,-2)}{4}{0}
  \midbar{blue!10}{(4,-2)}{1}{1}
  \midbar{blue!10}{(5,-2)}{1}{2}
  \lastbar{blue!10}{(6,-2)}{.96}{3}

  \node[left] at (0,-3) {\textit{numWrites}};
  \firstbar{yellow!10}{(0,-3)}{1}{0}
  \midbar{yellow!10}{(1,-3)}{.7}{1}
  \midbar{yellow!10}{(1.7,-3)}{.7}{2}
  \midbar{yellow!10}{(2.4,-3)}{2.6}{3}
  \lastbar{yellow!10}{(5,-3)}{1.96}{4}

  \node[left] at (0,-4) {\textit{safe}};
  \firstbar{green!50!black!10}{(0,-4)}{2.4}{\testrue}
  \midbar{red!10}{(2.4,-4)}{1.6}{\tesfalse}
  \lastbar{green!50!black!10}{(4,-4)}{2.96}{\testrue}
\end{tikzpicture}

  \end{minipage}
\end{center}
Read and write events occur independently at different frequencies.
The derived stream \emph{numReads} (\emph{numWrites}) counts the number of events of the input stream \emph{read} (\emph{write}).
While the \emph{read} and \emph{write} streams contain only discrete events, the number of events can be seen as a piece-wise constant signal with the initial value of 0.
The difference between the two signals is evaluated every time one of the two signals changes its value using the \emph{last known value} of both signals.
We call this concept \emph{signal semantics}: TeSSLa handles internally only streams of discrete events, but one can express operators following signal semantics in TeSSla and hence these discrete events can be seen as those points in time where the signal changes its value.
In these introductory examples operators are automatically lifted to signal semantics, which is formally introduced as the $\mathbf{slift}$ operator later.

\paragraph{Recursive Equations}
Like existing SRV approaches, TeSSLa relates a set of input streams to a set of output streams via mutually recursive equations, which allows self-references to the past, e.g. counting events of a stream \emph{x} as in the previous example is expressed in TeSSLa as follows:
$\mathit{count} := \mathbf{merge}(\mathbf{last}(\mathit{count}, x) + 1, 0)$
The $\mathbf{last}$ operator outputs the last known value of the \emph{count} stream, on every event of the stream \emph{x}.
The base of the recursion is provided by merging with 0, which is a stream with one initial event of value 0.
Since $\mathbf{last}$ only refers to events strictly lying in the past, the unique solution of such recursive equations can be computed incrementally (see \autoref{sec:semantics}). 

\paragraph{Time as First-Class Citizen}
In TeSSLa, every event has a timestamp which can be accessed via the $\mathbf{time}$ operator.
Since every event has a timestamp which is referring to a global clock and is unique for its stream, accessing the timestamps of events serves two purposes: Accessing the global order of events by comparing timestamps and performing calculations with the timestamps.
Consider e.g. the following specification which checks whether the lapse of time between two write events exceeds 5 time units and outputs the overtime if it does:
  
  \noindent
  \begin{minipage}{7cm}
    \setlength{\belowdisplayskip}{0pt}\setlength{\abovedisplayskip}{1ex}
    \begin{align*}
      \mathit{diff} &:= \mathbf{time}(\mathit{write}) - \mathbf{last}(\mathbf{time}(\mathit{write}), \mathit{write})\\
      \mathit{error} &:= \mathbf{filter}(\mathit{diff} > 5, \mathit{diff} - 5)
    \end{align*}
  \end{minipage}
  \quad
  \begin{minipage}{4.5cm}

\begin{tikzpicture}[events,xscale=.35]
  \node[unit, fill=blue!10, label=above:2] at (2,0) (a1) {};
  \node[unit, fill=blue!10, label=above:5] at (5,0) (a2) {};
  \node[unit, fill=blue!10, label=above:7] at (7,0) (a3) {};
  \node[unit, fill=blue!10, label=above:15] at (15,0) (a4) {};
  \node[unit, fill=blue!10, label=above:18] at (18,0) (a5) {};
  \path
    (0,0) node[left] {\textit{write}} edge[|-]
    (a1) (a1) edge (a2) (a2) edge (a3) (a3) edge (a4) (a4) edge (a5) (a5) edge[->] +(2,0);

  \node[event, fill=yellow!10] at (5,-1) (d1) {3};
  \node[event, fill=yellow!10] at (7,-1) (d2) {2};
  \node[event, fill=yellow!10] at (15,-1) (d3) {8};
  \node[event, fill=yellow!10] at (18,-1) (d4) {3};
  \path
    (0,-1) node[left] {\textit{diff}} edge[|-]
    (d1) (d1) edge (d2) (d2) edge (d3) (d3) edge (d4) (d4) edge[->] +(2,0);

  \node[event, fill=red!10] at (15,-2) (e1) {3};
  \path
    (0,-2) node[left] {\textit{error}} edge[|-]
    (e1) (e1) edge[->] +(5,0);
\end{tikzpicture}

  \end{minipage}

\vskip2pt

In the example, the stream $\mathit{diff} - 5$ is filtered by the condition $\mathit{diff} > 5$. Note that the property violation is only reported when the delayed event happens. To report such errors as soon as possible, TeSSLa has the ability to create events at certain points in time via the $\mathbf{delay}$ operator. The following specification checks the same property but raises a unit event on the \textit{error} stream as soon as we know that there was no \textit{write} event in time:

  \begin{minipage}{6cm}
    \setlength{\belowdisplayskip}{0pt}\setlength{\abovedisplayskip}{1ex}
    \begin{align*}
      \mathit{timeout} &:= \mathbf{const}(5)(\mathit{write})\\
      \mathit{error} &:= \mathbf{delay}(\mathit{timeout}, \mathit{write})
    \end{align*}
  \end{minipage}
  \begin{minipage}{5cm}

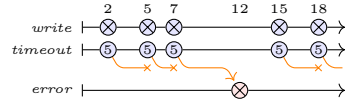
\begin{tikzpicture}[events,xscale=.35]
  \foreach \t in {2,5,7,12,15,18}
    \node at (\t,.8) {\t};

  \node[unit, fill=blue!10] at (2,0) (a1) {};
  \node[unit, fill=blue!10] at (5,0) (a2) {};
  \node[unit, fill=blue!10] at (7,0) (a3) {};
  \node[unit, fill=blue!10] at (15,0) (a4) {};
  \node[unit, fill=blue!10] at (18,0) (a5) {};
  \path
    (0,0) node[left] {\textit{write}} edge[|-]
    (a1) (a1) edge (a2) (a2) edge (a3) (a3) edge (a4) (a4) edge (a5) (a5) edge[->] +(2,0);

  \node[event, fill=blue!10] at (2,-1) (t1) {5};
  \node[event, fill=blue!10] at (5,-1) (t2) {5};
  \node[event, fill=blue!10] at (7,-1) (t3) {5};
  \node[event, fill=blue!10] at (15,-1) (t4) {5};
  \node[event, fill=blue!10] at (18,-1) (t5) {5};
  \path
    (0,-1) node[left] {\textit{timeout}} edge[|-]
    (t1) (t1) edge (t2) (t2) edge (t3) (t3) edge (t4) (t4) edge (t5) (t5) edge[->] +(2,0);

  \node[unit, fill=red!10] at (12,-2.8) (e1) {};
  \path
    (0,-2.8) node[left] {\textit{error}} edge[|-]
    (e1) (e1) edge[->] +(8,0);

  \draw[orange,-{Rays[length=3.5pt]}, rounded corners=2pt, shorten >=-1.75pt]
    (t1) -- ++(.8,-.8) -- ++(2.1,0);
  \draw[orange,-{Rays[length=3.5pt]}, rounded corners=2pt, shorten >=-1.75pt]
    (t2) -- ++(.8,-.8) -- ++(1.1,0);
  \draw[orange,->, shorten >=1pt, rounded corners=2pt]
    (t3) -- ++(.8,-.8) -- ++(3.2,0) -- (e1);
  \draw[orange,-{Rays[length=3.5pt]}, rounded corners=2pt, shorten >=-1.75pt]
    (t4) -- ++(.8,-.8) -- ++(2.1,0);
  \draw[orange, rounded corners=2.5pt]
    (t5) -- ++(.8,-.8) -- ++(1,0);
\end{tikzpicture}

  \end{minipage}

\vskip2pt

The $\mathbf{delay}$ function works as a timer, which is set to a timeout value with the first argument and reset with any event on the second argument. In the example, the function $\mathbf{const}(5)(\mathit{write})$ maps the values of events to the constant value of 5, which is then used as timeout value.
While in all the other examples the derived streams only contain events with timestamps taken from the input streams, in this example events with additional timestamps are generated. Like $\mathbf{last}$, the $\mathbf{delay}$ operator can be used in recursive equations, for example the equation
\[ \mathit{period} := \mathbf{merge}(\mathbf{const}(5)(\mathbf{delay}(\mathit{period}, \mathbf{unit})), 5) \]
produces an infinite stream with an event every 5 time units. The $\mathbf{merge}$ is used to provide a base case for the recursion and $\mathbf{const}$ is used to map the value of the generated events to 5 so that they can be used as the new timeout value.

\paragraph{Efficient Parallel Evaluation}
TeSSLa's design follows two principles to allow efficient evaluation on parallel hardware: \emph{Explicit memory usage} and \emph{local operator composition}.
If TeSSLa operates only on streams with bounded data-types of constant size, then the operators only need finite memory because every operator only needs to store at most one data value.
This allows implementations on systems without random access memory, e.g. FPGAs or embedded systems.
TeSSLa consists of a small set of primitive operators which can be flexibly combined.
The TeSSLa semantics is defined in a way that allows a local composition of the individual operators, which can be realized via message passing without the need for global synchronization.
Because of an explicit notion of progress for every stream describing how far the stream is known, local message passing is also sufficient to compute solutions for the recursive TeSSLa equations.
Implementing an efficient evaluation on FPGAs is part of our EU research project COEMS\footnote{\url{https://www.coems.eu}}.

\subsubsection{Related Work and Comparison}

LOLA~\cite{LOLA,LOLA2} is a synchronous stream specification language in the following sense: Events arrive in discrete steps and for every step, all input streams provide an event and all output streams produce an event, which means that it is not suitable for handling events with arbitrary real-time timestamps arriving at variable frequencies.
The not yet formally published RTLola~\cite{rtlola} is an extension of LOLA which introduces asynchronous streams to perform aggregations over real-time intervals. A major difference between RTLola and TeSSLa is that RTLola focuses on splitting input streams and aggregating over them, whereas TeSSLa provides a more general framework that in particular allows the (recursive) definition of aggregation operators while giving strict memory guarantees at the same time. 
Focus~\cite{DBLP:series/mcs/BroyS01} is a formalism for the specification of stream-based systems. Their timed streams progress by discrete ticks that separate events inbetween, thereby allowing multiple events at the same timestamp.
The synchronous stream programming languages Lustre~\cite{halbwachs87lustre}, Esterel~\cite{berry00foundations} and Signal~\cite{gautier87signal}, the stream specification language Copilot~\cite{pike10copilot} as well as the class of functional reactive programming (FRP) languages~\cite{eliot97functional} allow the description of the transformation in a linear style, i.e. an input stream is read chronologically and is thereby evaluated.
TeSSLa also supports linear evaluation because there are no future-references and the number of past-references is limited by the specification size.
The only complement to linear evaluation is the creation of additional events via the $\mathbf{delay}$ operator.
Quantitative regular expressions (QREs)~\cite{QRE} and logics like Signal Temporal Logic (STL)~\cite{STL} and Time-Frequency Logic (TFL)~\cite{donze12ontemporal} allow the mapping from complete streams to one final verdict/quantity.
They cannot generally be evaluated in a linear way.
The idea used in TeSSLa of supporting signals and event streams has also been used for Timed Regular Expressions~\cite{TRE}, but those have two explicitly different stream types, where TeSSLa internally represents signals as event streams.
Recently, synthesis of hardware-based monitors from stream specifications has become an important field:
For LOLA~\cite{LOLA} constant memory bounds for an algorithm that evaluates well-formed specifications exist and for LOLA 2.0~\cite{LOLA2} future references must be eliminated to gain constant memory bounds.
There has been work on synthesis of STL to FPGAs in different ways as well~\cite{eziostl2fpga,ezio17runtime}.


\section{Formal Definition of the TeSSLa Core Language}
\label{sec:semantics}

\noindent
In this section we introduce syntax and semantics of the minimal core of TeSSLa. In examples we use parametrized definitions, e.g $\mathbf{merge}(x, y) := …$ on top, which are expanded to their definitions until only core operators remain.

\paragraph{Preliminaries}
Given a partial order $(A, \le)$, a set $D⊆A$ is called \emph{directed} if $\forall a, b \in D: a \le b \vee b \le a$.
$(A, \le)$ is called \emph{directed-complete partial order (dcpo)} if there exists a supremum $\bigvee D$ for every directed subset $D \subseteq A$.
Let $f \in A \rightarrow B$ be a function and $(A, \le)$, $(B, \le')$ partial orders.
$f$ is called \emph{monotonic} if it preserves the order, i.e. $\forall a_1, a_2 \in A: a_1 \le a_2 \Rightarrow f(a_1) \le' f(a_2)$.
$f$ is called \emph{continuous} if it preserves the supremum, i.e. $\bigvee f(D) = f(\bigvee' D)$ for all directed subsets $D \subseteq A$.
By the Kleene fixed-point theorem, every monotonic and continuous function $f: A \to A$ has a least fixed point $\mu f$ if $(A, \le)$ is a dcpo with a least element $⊥$. $\mu f$ is the least upper bound of the chain iterating $f$ starting with the bottom element: $\mu f = \bigvee \{ f^n(\bot) \mid n \in \mathbb N \}$.

\paragraph{Syntax}

A TeSSLa specification $\phi$ consists of a set of possibly mutually recursive stream definitions defined over a finite set of variables $𝕍$ where an equation has the form $x := e$ with $x∈𝕍$ and
$$e ::= \mathbf{nil} ~|~ \mathbf{unit} ~|~ x ~|~ \mathbf{lift}(f)(e, …, e) ~|~ \mathbf{time}(e)~|~ \mathbf{last}(e, e)~|~ \mathbf{delay}(e, e).$$
All variables not occuring on the left-hand side of equations are \emph{input variables}.
All variables on the left-hand side are \emph{output variables}.
We call a TeSSLa specification \emph{flat} if it does not contain any nested expressions.
Every specification can be represented as a flat specification by using additional variables and equations.

\paragraph{Semantics}
We define the semantics of TeSSLa in terms of an abstract time domain which only requires a total order and corresponding arithmetic operators:
\begin{definition}
A \emph{time domain} is a totally ordered semi-ring $(𝕋,0,1,+,·,≤)$ that is not negative, i.e. $∀_{t∈𝕋}\, 0 ≤ t$.
\end{definition}
\noindent We extend the order on time domains to the set $𝕋_∞ = 𝕋 ∪ ｛∞｝$ with $∀_{t∈𝕋}\, t < ∞$.

Conceptually, streams are timed words that are known inclusively or exclusively up to a certain timestamp, its progress, that might be infinite. A stream might contain an infinite number of events even if its progress is finite.
\begin{definition}
An \emph{event stream} over a time domain $𝕋$ and a data domain $𝔻$ is a finite or infinite sequence $s = a_0 a_1 \dots ∈ 𝓢_𝔻 = (𝕋·𝔻)^ω ∪ (𝕋·𝔻)^+ ∪ (𝕋·𝔻)^*·(𝕋_∞ ∪ 𝕋·｛\streamincl｝)$ where $a_{2i} < a_{2(i+1)}$ for all $i$ with $0<2(i+1)<|s|$ ($|s|$ is $∞$ for infinite streams). The \emph{prefix relation} over $𝓢_𝔻$ is the least relation that satisfies $s \sqsubseteq s$, $u \sqsubseteq s$ if $uv \sqsubseteq s$ and $ut'\streamincl \sqsubseteq s$ if $ut \sqsubseteq s$, $t' < t$, $t∈𝕋_∞$ and $t'∈𝕋$.
\end{definition}
We say a stream has an event with value $d$ at time $t$ if in its sequence $d$ directly follows $t$.
We say a stream is known at time $t$ if it contains a strictly larger timestamp or a non-strictly larger timestamp followed by a data value or $\bot$.
Where convenient, we also see streams as functions $s \in 𝕋→𝔻∪｛⊥,\mathrm{?}｝$ such that $s(t) = d$ if the stream has value $d$ at time $t$, $s(t) = ⊥$ if it is known to have no value, and $s(t) = \mathrm{?}$ otherwise.
We refer to the supremum of all known timestamps of a stream as inclusive or exclusive progress, depending on whether it is itself a known timestamp. 
The prefix relation realises the intuition of cutting a stream at a certain point in time while keeping or removing the cutting point.

In the following, we present the denotation of a specification $\phi$ as a function between input streams and output streams.

\begin{definition}[TeSSLa semantics]\label{def:tessla-semantics}
  Given a specification $\phi$ of equations $y_i := e_i$, every $e_i$ can be interpreted as a function of input streams $s_1, \ldots, s_k$ and output streams $s'_{1}, \ldots, s'_{n}$, that is composed of the primitive functions whose denotation is given in the rest of this section. Input variables are mapped to input streams, $\sem{x_i}_{s_1, \ldots, s_k, s'_{1}, \ldots, s'_{n}} = s_i$ and output variables to output streams, $\sem{y_i}_{s_1, \ldots, s_k, s'_{1}, \ldots, s'_{n}} = s'_i$. Thus for fixed input streams $s_1, \ldots, s_k$ and every $e_i$, we obtain a function $\sem{e_i}_{s_{1}, \ldots, s_{k}} \in \mathcal S_{\mathbb{D}'_{1}}\times \ldots \times \mathcal S_{\mathbb{D}'_{n}} \rightarrow \mathcal S_{\mathbb{D}'_{i}}$ and in combination a function $\sem{e_1,…,e_n}_{s_{1}, \ldots, s_{k}} \in \mathcal S_{\mathbb{D}'_{1}}\times \ldots \times \mathcal S_{\mathbb{D}'_{n}} \rightarrow \mathcal S_{\mathbb{D}'_{1}}\times \ldots \times \mathcal S_{\mathbb{D}'_{n}}$.
  We now define the denotation of a specification $\phi$ as the least fixed-point of this function.
  \begin{align*}
  \sem\phi &\in \mathcal S_{\mathbb{D}_{1}}\times \ldots \times \mathcal S_{\mathbb{D}_{k}} \rightarrow \mathcal S_{\mathbb{D}'_1}\times \ldots \times \mathcal S_{\mathbb{D}'_n}\\
  \sem\phi &(s_{1}, \ldots, s_{k}) = \mu \left( \sem {e_1,…,e_n}_{s_{1}, \ldots, s_{k}} \right) 
  \end{align*}
\end{definition}

The function $\sem{e_1,…,e_n}_{s_{1}, \ldots, s_{k}}$ is monotonic and continuous because all primitive TeSSLa functions defined later in this section are monotonic and continuous and both properties are closed under function composition and cartesian products.
$(𝓢_𝔻, \sqsubseteq)$ and by extension $(𝓢_{𝔻_1} \times \ldots \times 𝓢_{𝔻_n}, \sqsubseteq \times \ldots \sqsubseteq)$ are dcpos.
By the Kleene fixed-point theorem $\sem {e_1,…,e_n}_{s_{1}, \ldots, s_{k}}$ has a least fixed point, which is the least upper bound of its Kleene chain.

Next we give the semantics of the primitive TeSSLa functions. The dependency of the input and output streams $s_{1}, \ldots, s_{k}, s'_1, ..., s'_n$ is assumed implicitly.

\begin{definition}
\emph{Nil} is a constant for the completely known stream without any events: $\sem{\mathbf{nil}} = ∞ \in 𝓢_𝔻$.
\end{definition}

\noindent We use the unit type $𝕌=｛\unitsym｝$ for streams that can carry only the single value $\unitsym$.

\begin{definition}
\emph{Unit} is a constant for the completely known stream with a single unit event at timestamp zero: $\sem{\mathbf{unit}} = 0 \unitsym ∞ \in 𝓢_𝕌$
\end{definition}

The following functions are given by specifying two conditions: the first for positions where an output event occurs, and the second where no output event occurs.
Thereby the progress of the stream is defined indirectly as the position where the output can no longer be inferred from these conditions.

\begin{definition}
The \emph{time} operator returns the stream of the timestamps of another stream $\sem{\mathbf{time}(e)} = \mathsf{time}(\sem{e})$ where $\mathsf{time} \in 𝓢_𝔻 → 𝓢_𝕋$ is defined as $\mathsf{time}(s) = s'$ such that
 \[
    ∀_{t} s'(t) = t ⇔ s(t) ∈ 𝔻 \qquad
    ∀_{t} s'(t) = ⊥ ⇔ s(t) = ⊥.
  \]
\end{definition}

The lift operator lifts an $n$-ary function $f$ from values to streams. The notation $A_1 \times \ldots \times A_n \rightarrowtail B$ denotes the set of functions where all $A_i$ and $B$ have been extended by the value $⊥$.

\begin{definition}
Unary \emph{lift} is defined as $\sem{\mathbf{lift}(f)(e)} = \mathsf{lift_1}(f)(\sem{e})$ where $\mathsf{lift}_1 \in (𝔻 \rightarrowtail 𝔻') → (𝓢_{𝔻}→𝓢_{𝔻'})$ is given by $\mathsf{lift}_1(f)(s) = s'$ such that
  \begin{align*}
    ∀_{t,d∈𝔻'} s'(t) = d &⇔ s(t) ∈ 𝔻 ∧ f(s(t)) = d\\
    ∀_{t} s'(t) = ⊥ &⇔ s(t) = ⊥ ∨ s(t) ∈ 𝔻 ∧ f(s(t)) = ⊥.
  \end{align*}
\end{definition}

\begin{definition}  
Binary \emph{lift} is given as $\sem{\mathbf{lift}(f)(e_1, e_2)} = \mathsf{lift_2}(f)(\sem{e_1}, \sem{e_2})$ where $\mathsf{lift}_2 \in (𝔻_1 × 𝔻_2 \rightarrowtail 𝔻') → (𝓢_{𝔻_1}×𝓢_{𝔻_2}→𝓢_{𝔻'})$ is given by $\mathsf{lift}_2(f)(s,s') = s''$ s.t.  
  \begin{align*}
    ∀_{t,d∈𝔻'} s''(t) = d &⇔ (s(t) ∈ 𝔻_1 ∨ s'(t) ∈ 𝔻_2) ∧ \mathsf{known}(t) ∧ f(s(t), s'(t)) = d\\
    ∀_{t} s''(t) = ⊥ &⇔ (s(t) = ⊥ ∧ s'(t) = ⊥) ∨ \mathsf{known}(t) ∧ f(s(t), s'(t)) = ⊥
  \end{align*}
  where $\mathsf{known}(t) := s(t) ≠ \mathrm{?} ∧ s'(t) ≠ \mathrm{?}$.
\end{definition}

The binary lift can naturally be extended to an $n$-ary lift by recursively combining two streams into a stream of tuples or partially applied functions until the final result is obtained (see Appendix~\ref{def:n-ary-lift}). Alternatively, the scheme of the binary lift can be easily extended to higher arities.

\begin{example}
  \emph{Merge} combines events of two streams, prioritising the first one.

  \begin{minipage}{6.5cm}
    \setlength{\belowdisplayskip}{0pt}\setlength{\abovedisplayskip}{1ex}
    \[\mathbf{merge}(x, y) := \mathbf{lift}(\mathsf{mergeaux})(x, y)\]
    \begin{align*}
      \mathsf{mergeaux}(a≠⊥, b) &:= a\\
      \mathsf{mergeaux}(⊥, b) &:= b
    \end{align*}
  \end{minipage}
  \begin{minipage}{4cm}
    \begin{tikzpicture}[events]
      \node[event, fill=red!10] at (1,0) (x1) {$6$};
      \node[event, fill=red!10] at (3,0) (x2) {$4$};
      \node[event, fill=red!10] at (4,0) (x3) {$2$};

      \node[event, fill=blue!10] at (2,-1) (y1) {$5$};
      \node[event, fill=blue!10] at (4,-1) (y2) {$7$};

      \node[event, fill=red!10] at (1,-2) (z1) {$6$};
      \node[event, fill=blue!10] at (2,-2) (z2) {$5$};
      \node[event, fill=red!10] at (3,-2) (z3) {$4$};
      \node[event, fill=red!10] at (4,-2) (z4) {$2$};

      \path
        (0,0) node[left] {$x$} edge[|-] (x1)
        (x1) edge (x2)
        (x2) edge (x3)
        (x3) edge[->] (5,0);
      \path
        (0,-1) node[left] {$y$} edge[|-] (y1)
        (y1) edge (y2)
        (y2) edge[->] (5,-1);
      \path
        (0,-2) node[left] {$\mathbf{merge}(x,y)$} edge[|-] (z1)
        (z1) edge (z2)
        (z2) edge (z3)
        (z3) edge (z4)
        (z4) edge[->] (5,-2);
    \end{tikzpicture}
  \end{minipage}
\end{example}

\begin{example}
  \emph{Const} maps the values of all events of the input stream to a constant value: $\mathbf{const}(c)(a) := \mathbf{lift}(\mathsf{constaux}(c))(a)$ with $\mathsf{constaux}(c)(a) := c$.
  Using $\mathbf{const}$ we can lift constants into streams representing a constant signal with this value, e.g. $\mathbf{true} := \mathbf{const}(\op{true})(\mathbf{unit})$ or $\mathbf{zero} := \mathbf{const}(0)(\mathbf{unit})$.
\end{example}

\begin{definition}
  The \emph{last} operator takes two streams and returns the previous value of the first stream at the timestamps of the second.
  It is defined as $\sem{\mathbf{last}(e_1, e_2)} = \mathsf{last}(\sem{e_1}, \sem{e_2})$ where $\mathsf{last}_{𝔻,𝔻'} \in 𝓢_𝔻 × 𝓢_{𝔻'} → 𝓢_𝔻$ is given as $\mathsf{last}(s, s') = s''$ such that
  \begin{align*}
    ∀_{t,d∈𝔻} s''(t) = d &⇔ s'(t) ∈ 𝔻' ∧ ∃_{t' < t} s(t') = d ∧ \mathsf{noData}(t', t)\\
    ∀_{t} s''(t) = ⊥ &⇔ s'(t) = ⊥ ∧ \mathsf{defined}(t) ∨ ∀_{t' < t} s(t') = ⊥
  \end{align*}
  where $\mathsf{noData}(t, t') := ∀_{t'' | t < t'' < t'} s(t'') = ⊥$ and $\mathsf{defined}(t) := \forall_{t' < t} s''(t') ≠ \mathrm{?}$.
\end{definition}

\noindent Note that while TeSSLa is defined on event streams, \textbf{last} realizes some essential aspects of the signal semantics: With this operator one can query the last known value of an event stream at a specific time and hence interpret the events on this stream as points where a piece-wise constant signal changes its value.

\begin{example}
  By combining the $\mathbf{last}$ and the $\mathbf{lift}$ operators, we can now realize the \emph{signal lift} semantics implicitly used in the introduction:\\ $\mathbf{slift}(f)(x, y) := \mathbf{lift}(\mathsf{sliftaux}(f))(x',y')$ with

  \begin{minipage}{6.5cm}
    \setlength{\belowdisplayskip}{0pt}\setlength{\abovedisplayskip}{1ex}
    \begin{align*}
      x' &:= \mathbf{merge}(x, \mathbf{last}(x, y)) \text{ and}\\
      y' &:= \mathbf{merge}(y, \mathbf{last}(y, x)).
    \end{align*}
    \begin{align*}
      \mathsf{sliftaux}(f)(a≠⊥, b≠⊥) &:= f(a, b)\\
      \mathsf{sliftaux}(f)(⊥, b) &:= ⊥\\
      \mathsf{sliftaux}(f)(a, ⊥) &:= ⊥
    \end{align*}
  \end{minipage}
  \begin{minipage}{4cm}
    \begin{tikzpicture}[events]
      \node[event, fill=red!10] at (1,0) (x1) {$1$};
      \node[event, fill=red!10] at (3,0) (x2) {$5$};
      \node[event, fill=red!10] at (4,0) (x3) {$3$};
      \node[event, fill=red!10] at (5,0) (x4) {$1$};

      \node[event, fill=blue!10] at (2,-1) (y1) {$2$};
      \node[event, fill=blue!10] at (5,-1) (y2) {$4$};

      \node[event, fill=red!10] at (1,-2) (x'1) {$1$};
      \node[event, fill=red!10] at (2,-2) (x'2) {$1$};
      \node[event, fill=red!10] at (3,-2) (x'3) {$5$};
      \node[event, fill=red!10] at (4,-2) (x'4) {$3$};
      \node[event, fill=red!10] at (5,-2) (x'5) {$1$};

      \node[event, fill=blue!10] at (2,-3) (y'1) {$2$};
      \node[event, fill=blue!10] at (3,-3) (y'2) {$2$};
      \node[event, fill=blue!10] at (4,-3) (y'3) {$2$};
      \node[event, fill=blue!10] at (5,-3) (y'4) {$4$};

      \node[event, fill=yellow!10] at (2,-4) (z1) {$3$};
      \node[event, fill=yellow!10] at (3,-4) (z2) {$7$};
      \node[event, fill=yellow!10] at (4,-4) (z3) {$5$};
      \node[event, fill=yellow!10] at (5,-4) (z4) {$5$};

      \path
        (0,0) node[left] {$x$} edge[|-]
        (x1) (x1) edge (x2) (x2) edge (x3) (x3) edge (x4) (x4) edge[->] +(1,0);
      \path
        (0,-1) node[left] {$y$} edge[|-]
        (y1) (y1) edge (y2) (y2) edge[->] +(1,0);
      \path
        (0,-2) node[left] {$x'$} edge[|-]
        (x'1) (x'1) edge (x'2) (x'2) edge (x'3) (x'3) edge (x'4) (x'4) edge (x'5) (x'5) edge[->] +(1,0);
      \path
        (0,-3) node[left] {$y'$} edge[|-]
        (y'1) (y'1) edge (y'2) (y'2) edge (y'3) (y'3) edge (y'4) (y'4) edge[->] +(1,0);
      \path
        (0,-4) node[left] {$x + y$} edge[|-]
        (z1) (z1) edge (z2) (z2) edge (z3) (z3) edge (z4) (z4) edge[->] +(1,0);
    \end{tikzpicture}
  \end{minipage}
\end{example}

\begin{example}
  In order to \emph{filter} an event stream with a dynamic condition, we apply the last known filter condition to the current event:\\
  $\mathbf{filter}(z, x) := \mathbf{lift}(\mathsf{filteraux})(\mathbf{merge}(z, \mathbf{last}(z, x)), x)$

  \begin{minipage}{6.5cm}
    \setlength{\belowdisplayskip}{0pt}\setlength{\abovedisplayskip}{1ex}
    \begin{align*}
      \mathsf{filteraux}: 𝔹 × A &\rightarrowtail A\\
      \mathsf{filteraux}(c \neq \op{true}, a) &= \bot\\
      \mathsf{filteraux}(\op{true}, a) &= a
    \end{align*}
  \end{minipage}
  \begin{minipage}{4cm}
    \begin{tikzpicture}[events]
      \node[event, fill=green!50!black!10] at (2,0) (z1) {\testrue};
      \node[event, fill=red!10] at (4,0) (z2) {\tesfalse};

      \node[unit, fill=blue!10] at (1,-1) (x1) {};
      \node[unit, fill=blue!10] at (3,-1) (x2) {};
      \node[unit, fill=blue!10] at (4,-1) (x3) {};
      \node[unit, fill=blue!10] at (5,-1) (x4) {};

      \node[event, fill=green!50!black!10] at (2,-2) (z'1) {\testrue};
      \node[event, fill=green!50!black!10] at (3,-2) (z'2) {\testrue};
      \node[event, fill=red!10] at (4,-2) (z'3) {\tesfalse};
      \node[event, fill=red!10] at (5,-2) (z'4) {\tesfalse};

      \node[unit, fill=blue!10] at (3,-3) (f1) {};

      \path
        (0,0) node[left] {$z$} edge[|-]
        (z1) (z1) edge (z2) (z2) edge[->] +(2,0);
      \path
        (0,-1) node[left] {$x$} edge[|-]
        (x1) (x1) edge (x2) (x2) edge (x3) (x3) edge (x4) (x4) edge[->] +(1,0);
      \path
        (0,-2) node[left] {$z'$} edge[|-]
        (z'1) (z'1) edge (z'2) (z'2) edge (z'3) (z'3) edge (z'4) (z'4) edge[->] +(1,0);
      \path
        (0,-3) node[left] {$\mathbf{filter}(z, x)$} edge[|-]
        (f1) (f1) edge[->] +(3,0);
    \end{tikzpicture}
  \end{minipage}
\end{example}

\begin{definition}
The \emph{delay} operator takes delays as its first argument. After a delay has passed, a unit event is emitted.
A delay can only be set if a reset event is received via the second argument, or if an event is emitted on the output.
Formally, $\sem{\mathbf{delay}(e_1, e_2)} = \mathsf{delay}(\sem{e_1}, \sem{e_2})$ where $\mathsf{delay}_{𝔻} \in 𝓢_{𝕋∖｛0｝}×𝓢_𝔻 → 𝓢_𝕌$ is given as $\mathsf{delay}(s, s') = s''$ such that
\begin{align*}
∀_t s''(t) = \boxempty &⇔ ∃_{t' < t} s(t') = t - t' ∧ \mathsf{setable}(t') ∧ \mathsf{noreset}(t', t)\\
∀_t s''(t) = ⊥ &⇔ \mathsf{defined}(t) ∧ ∀_{t' < t} s(t') ≠ \mathrm{?} ∧ s(t') ≠ t - t' ∨ \mathsf{unsetable}(t') ∨ \mathsf{reset}(t', t)
\end{align*}
where $\mathsf{setable}(t) := s''(t) = \boxempty ∨ s'(t) ∈ 𝔻$, $\mathsf{unsetable}(t) := s''(t) = ⊥ ∧ s'(t) = ⊥$, $\mathsf{noreset}(t, t') := ∀_{t'' | t < t'' < t'} s'(t'') = ⊥$ and $\mathsf{reset}(t, t') := ∃_{t'' | t < t'' < t'} s'(t'') ∈ 𝔻$.
\end{definition}
In many applications the delay operator is used in simplified versions: In the first example of the introduction that uses the delay operator, the delay and the reset argument can be the same because the delay is used only in non-recursive equations and every new delay is a reset, too.
If a periodical event pattern is generated independently from input events then the second argument can be set to unit because only an initial reset event is needed.
The full complexity of the delay operator is only needed if the delay is used in recursive equations with input dependencies and ensures that the fixed-point is unique.

We can observe that all basic functions are monotonic and continuous.
From the fact, that these properties are closed under composition and the smallest fixed-point is determined by the Kleene chain, we can therefore conclude:
\begin{proposition}\label{thm:semmonotonicity}
The semantics of a TeSSLa specification is monotonic and continuous in the input streams.
\end{proposition}
\noindent In other words, the semantics will provide an extended result for an extended input and is therefore suited for online monitoring.

We can further observe that the pre-fixed-points on the Kleene chain have the following property: the progress only increases a finite number of times until a further event has to be appended.
This is due to the basic functions that do handle progress in this way.
We therefore obtain:
\begin{theorem}
  For a specification $φ$ every finite prefix of $\sem{φ}(s_1, \ldots, s_k)$ can be computed assuming all lifted functions are computable.
  Assuming they are computable in $O(1)$ steps, the prefix can be computed in $O(k·|φ|)$ steps where $k$ is the number of events over all involved streams.
\end{theorem}
Note that in case the specification contains no $\mathbf{delay}$ output streams cannot contain any such timestamps that did not occur already in the inputs.
Further note, that fixed-points might contain infinitely many positions with data values (in case of $\mathbf{delay}$) and we can thus only compute prefixes.
A respective monitor would exhibit infinite outputs even for finite inputs.

Due to \autoref{thm:semmonotonicity} we can reuse a previously computed fixed-point if new input events occur and hence also compute the outputs incrementally.

\subsubsection{Well-formedness}
While the least fixed-point is unique it does not have to be the only fixed-point.
In that case, the least fixed-point is often the stream with progress $0$ or some other stream with too little progress and one would be interested in (one of) the maximal fixed-points.
Since the largest fixed-points would be more difficult to compute, especially in the setting of online monitoring, we define a fragment for which a unique fixed-point exists.

\begin{definition}
  We call a TeSSLa specification $\phi$ \emph{well-formed} if every cycle of the dependency graph (of the flattened specification) contains at least one \emph{delayed}-labelled edge.
  The \emph{dependency graph} of a flat TeSSLa specification $\phi$ of equations $y_i := e_i$ is the directed multi-graph $G = (V,E)$ of nodes $V = \{y_1, \ldots, y_n\}$. For every $y_i := e_i$ the graph contains the edge $(y_i, y_j)$ iff $y_j$ is used in $e_i$. We label edges corresponding to the first argument of $\mathbf{last}$ or $\mathbf{delay}$ with \emph{delayed}.
\end{definition}

\begin{theorem}\label{thm:unique-fixed-point}
  Given a well-formed specification $\phi$ of equations $y_i := e_i$ and input streams $s_1, \ldots, s_k$ then $\mu (\sem{e_1, \ldots, e_n}_{s_{1}, \ldots, s_{k}})$ is the only fixed-point.
\end{theorem}

\begin{proof}
  From the Kleene fixed-point theorem we know
  $\mu \left( \sem {e_1, \ldots, e_n}_{s_{1}, \ldots, s_{k}} \right) =$\\
  $ \bigsqcup \{ \sem {e_1, \ldots, e_n}^n_{s_{1}, \ldots, s_{k}}(\bot) \mid n \in \mathbb N \}$.
  Because $\phi$ is well-formed, every $\sem{e_i}_{s_{1}, \ldots, s_{k}}$ is either constant or contains at least one $\mathbf{last}$ or $\mathbf{delay}$. The input streams $s_{1}, \ldots, s_{k}$ limit progress, i.e. the maximal timestamp produced, of $\sem{e_i}_{s_{1}, \ldots, s_{k}}$. The progress strictly increases with every step of the iteration of $\sem{e_1,\ldots,e_n}_{s_{1}, \ldots, s_{k}}$ in the Kleene chain until the limit given by the input streams is reached. Every other fixed-point of $\sem{e_1,\ldots,e_n}_{s_{1}, \ldots, s_{k}}$ must be an extension of the least fixed-point, but the least fixed-point has already the maximal progress permitted by the input streams.
\end{proof}


\section{Expressiveness of TeSSLa}
\label{sec:properties}

We discuss the expressiveness of four different TeSSLa fragments: TeSSLa specifications without the delay operator can only produce events with timestamps which are already included in the input streams and TeSSLa specifications with the delay operator can produce arbitrary event patterns even without any input event. On the other hand we distinguish between TeSSLa specifications which use only bounded data structures, which can only consider finitely many past events, and those with unbounded data structures which can consider infinitely many past events in the computation of new events.

To characterize functions which can be expressed in TeSSLa we define \emph{timestamp conservatism} and \emph{future independence} in addition to monotonicity and continuity. For a stream $a \in \mathcal S_{\mathbb D}$ we denote with $T(a)$ the set of timestamps present in the stream $a$ and for multiple streams $T(a_1, \ldots, a_n) := \bigcup_{1 \le i \le n} T(a_i)$.

\begin{definition}[Timestamp Conservatism]
  We call a function $f \in \mathcal S_{\mathbb D_1} \times \ldots \times \mathcal S_{\mathbb D_k} \to \mathcal S_{\mathbb D'_1} \times \ldots \times \mathcal S_{\mathbb D'_n}$ on streams \emph{timestamp conservative} iff it does not introduce new timestamps, i.e. for input streams $a \in \mathcal S_{\mathbb D_1} \times \ldots \times \mathcal S_{\mathbb D_k}$ and output streams $b \in \mathcal S_{\mathbb D'_1} \times \ldots \times \mathcal S_{\mathbb D'_n}$ we have $f(a) = b$ implies $T(a) \supseteq T(b)$.
\end{definition}

Note that TeSSLa specifications without delay are timestamp conservative because only delay can introduce new timestamps.

For a stream $a \in \mathcal S_{\mathbb D}$ we denote with $a|_t$ the prefix of $a$ with progress $t$.

\begin{definition}[Future Independence]
  We call a function $f \in \mathcal S_{\mathbb D_1} \times \ldots \times \mathcal S_{\mathbb D_k} \to \mathcal S_{\mathbb D'_1} \times \ldots \times \mathcal S_{\mathbb D'_n}$ on streams \emph{future independent} iff output events only depend on current or previous events, i.e. for input streams $a \in \mathcal S_{\mathbb D_1} \times \ldots \times \mathcal S_{\mathbb D_k}$ and output streams $b \in \mathcal S_{\mathbb D'_1} \times \ldots \times \mathcal S_{\mathbb D'_n}$ we have $f(a) = b$ implies $\forall_{t \in \mathbb T}\ f(a_1|_t, \ldots, a_k|_t) = (b_1|_t, \ldots, b_n|_t)$.
\end{definition}

Note that every TeSSLa specification is future independent because the operators $\mathbf{last}$ and $\mathbf{delay}$ are the only operators referring to events with different timestamps and they refer only to previous events.
The omitted proofs of the following theorems can be found in Appendix~\ref{sec:expressiveness-proofs}.

\begin{theorem}[Expressiveness of TeSSLa Without Delay]\label{thm:expressiveness-no-delay}
  Every function $f \in \mathcal S_{\mathbb D_1} \times \ldots \times \mathcal S_{\mathbb D_k} \to \mathcal S_{\mathbb D'_1} \times \ldots \times \mathcal S_{\mathbb D'_n}$ on streams can be represented as a TeSSLa specification without delay iff it is
  \begin{inparaenum}[a)]
    \item monotonic and continuous,
    \item timestamp conservative and
    \item future independent.
  \end{inparaenum}
\end{theorem}

\begin{proofsketch}
  Represent the function $f$ as the iterative function $\tilde f(m, d, t) = m'$ taking a memory state $m$, the current input values $d$, and the corresponding current timestamp $t$ and returning the new memory state $m'$.
  Output events for all output streams can be derived from $m'$.
  Because $f$ is monotonic it is sufficient to compute the output events step by step;
  because $f$ is future independent it is sufficient to allow $\tilde f$ to store arbitrary information about the past events;
  and because $f$ is timestamp conservative it is sufficient to execute $\tilde f$ for every timestamp in the input events.
  Translate $f(x_1, \ldots, x_k) = y_1, \ldots, y_n$ into an equivalent TeSSLa specification:
  $t := \mathbf{time}(\mathbf{merge}(x_1, \ldots, x_k)),\ 
  m := \mathbf{lift}(\tilde f)(\mathbf{last}(m, t), x_1, \ldots, x_k, t) \text{ and }
  \forall_{i \le n}\ y_i := \mathbf{lift}(\tilde o_i)(m)$.
\end{proofsketch}

If all data types in the TeSSLa specification $\phi$ are bounded, $\tilde f$ uses a finite memory cell $m$, which can only store a constant number of current and previous events.
Monotonicity guarantees that we can compute output events incrementally and by future independence we know that knowledge about the previous events is sufficient to derive new events.
From the combination of both properties we know that it is not necessary to queue (arbitrarily large) event sequences to compute the output events.
Instead one memory cell (capable of storing one element of the data domain) per delay and per last operator in the specification is sufficient.
Restricting TeSSLa to bounded data types allows TeSSLa implementations on embedded systems without addressable memory because then finite memory is sufficient. Such a restricted TeSSLa specification can compute new events only based on a finite number of current and previous events.

\begin{theorem}[Expressiveness of TeSSLa With Delay]\label{thm:expressiveness-delay}
  Every function $f \in \mathcal S_{\mathbb D_1} \times \ldots \times \mathcal S_{\mathbb D_k} \to \mathcal S_{\mathbb D'_1} \times \ldots \times \mathcal S_{\mathbb D'_n}$ can be represented as a TeSSLa specification with delay iff it is
  \begin{inparaenum}[a)]
    \item monotonic and continuous and
    \item future independent.
  \end{inparaenum}
\end{theorem}

The proof accompanies the step-function $\tilde f$ with a timeout function $\tilde u$ which is evaluated on every new memory state.
$\tilde u$ returns the timestamp of the next evaluation of $\tilde f$, which allows arbitrary event generation.
The effect of $\tilde u$ can be realized using the \textbf{delay} operator.

\noindent We call a stream \emph{Zeno} if it contains two timestamps $t_1$ and $t_2$ with infinitely many events between $t_1$ and $t_2$.
With the delay operator it is possible to construct such Zeno streams because the timeout function is not restricted in any way.
By Rice's theorem it is impossible to check for an arbitrary timeout function whether it only generates non-Zeno timestamp sequences.
Hence, one would need to restrict allowed timeout functions more drastically, which would restrict the possible event sequences generated by a TeSSLa specification further than necessary. For that reason we decided to include the capability to generate Zeno streams with TeSSLa.
As a consequence of \autoref{thm:expressiveness-delay} we obtain:
\begin{corollary}
  A TeSSLa specification with multiple delays can be translated into an equivalent specification with only one delay.
\end{corollary}

TeSSLa with and without delay are closely related because TeSSLa without delay can verify the relation of given input/output streams with respect to a TeSSLa specification that uses delay. The delay is only needed to actively generate the events at specified times.
In the following we denote with $\sem{\phi |_y}(x_1, \ldots, x_k) \in \mathbb B$ the boolean function indicating whether the boolean output stream $y \in \mathcal S_{\mathbb B}$ of the TeSSLa specification $\phi$ contains only events with value true for the input streams $x_1, \ldots, x_k \in \mathcal S_{\mathbb D_1} \times \ldots \times \mathcal S_{\mathbb D_k}$.

\begin{theorem}[Delay Elimination]
  For every TeSSLa specification $\phi$ with $\sem\phi \in \mathcal S_{\mathbb D_1} \times \ldots \times \mathcal S_{\mathbb D_k} \to \mathcal S_{\mathbb D'_1} \times \ldots \times \mathcal S_{\mathbb D'_n}$ with delay operators there exists a TeSSLa specification $\phi'$ without delay operators, which derives a boolean stream $z \in \mathcal S_{\mathbb B}$, s.t. for any input streams $x_1, \ldots, x_k$ and output streams $y_1, \ldots, y_n$ we have $\sem\phi(x_1, \ldots, x_k) = y_1, \ldots, y_n$ iff $\sem{\phi' |_z}(x_1, \ldots, x_k, y_1, \ldots, y_n)$.
\end{theorem}

The above theorem follows from \autoref{thm:expressiveness-no-delay} and the fact that $\sem{\phi' |_z}$ is timestamp conservative, because the output stream $z$ only contain events when any input stream contains an event. See Appendix~\ref{sec:delay-elimination-proof} for a constructive proof of a slightly weaker lemma.


\section{TeSSLa Fragments and Transducers}
\label{sec:transducers}

In this section we investigate two TeSSLa fragments related to deterministic Büchi automata and timed automata, resp.
We translate TeSSLa specifications to transducers, which can be seen as automata taking the in- and output of the corresponding transducer as input word.
Thus by relating TeSSLa fragments to certain transducer classes, we inherit complexity and expressiveness results from the well-known automata models.

\subsubsection{Boolean Fragment}

The fragment TeSSLa$_\text{bool}$ restricts TeSSLa to boolean streams and the operators \textbf{last}, \textbf{lift} and \textbf{slift} with $\geq$ on timestamps. In the syntax expressions are restricted as follows, where $f$ is a function $f: \mathbb{B}^n \rightarrowtail \mathbb{B}$:
\[e := \mathbf{nil} \mid \mathbf{unit} \mid x \mid \mathbf{lift}(f)(e,\dots,e) \mid \mathbf{slift}(\geq)(\mathbf{time}(e), \mathbf{time}(e)) \mid \mathbf{last}(e,e)\]
Note that since one can only compare timestamps, for a TeSSLa$_\text{bool}$-formula $\phi$ and two tuples of input streams $S, S' \in \mathcal S_{\mathbb D_1} \times \ldots \mathcal S_{\mathbb D_n}$ we have $\sem\phi(S) = \sem\phi(S')$ iff all events in $S'$ carry the same values in the same order as those in $S$, independent from the exact timestamps of the events.

A \emph{deterministic finite state transducer (DFST)} is a 5-tuple $R = (\Sigma, \Gamma, Q, q_0, \delta)$ with input alphabet $\Sigma$, output alphabet $\Gamma$, state set $Q$, initial state $q_0 \in Q$ and transition function $\delta: Q \times \Sigma \rightarrow Q \times \Gamma$.
For an input word $w = w_0w_1w_2\dots$ we call a sequence
$s_0 \xrightarrow{w_0 / o_0} s_1 \xrightarrow{w_1 / o_1} s_2 \xrightarrow{w_2 / o_2} \cdots$
a run of a DFST $R$ with output $\sem R(w) = o_0o_1o_2\dots \in \Gamma^\infty$ iff $s_0 = q_0$ and $\delta(s_i,w_i) = (s_{i+1},o_i)$ for all $i \geq 0$.
To show that TeSSLa$_\text{bool}$ and DFSTs have the same expressiveness, we encode DFST words as TeSSLa$_\text{bool}$ streams and vice versa.
The function $\alpha_\Sigma(w) = S$ encodes a DFST word $w = w_0w_1\dots \in \Sigma^\infty$ as a corresponding set of TeSSLa$_\text{bool}$ streams: For every $p \in \Sigma$ a stream $s_p \in S$ exists with
$s_p = 0d_01d_1\dots\infty \Leftrightarrow \forall i: (d_i \Leftrightarrow w_i = p)$.
The function $\beta_\Sigma(s_1, \ldots, s_k) = w = w_0w_1\dots \in \Sigma^\infty$ encodes TeSSLa$_\text{bool}$ streams as a synchronized DFST word $w$ over the alphabet $\Sigma = \{z_1, \ldots, z_k\} \to \op{Val}$ with $\op{Val} = \{\bot, d, <', \bot', d' \mid d \in \{\testrue, \tesfalse\}\}$ which maps stream names to their current values: Let $T = \{t_0 = 0, t_1, t_2, \ldots\}$ be the set of all timestamps present in the streams including $0$ with $t_i < t_{i+1}$. Then $w_i(s) = \op{<}'$ if $s$ has exclusive progress of $t_i$, $w_i(s) = s(t_i)'$ if $s$ has inclusive progress of $t_i$ or $w_i(s) = s(t_i)$ otherwise.
\begin{theorem}
  For a DFST $R = (\Sigma, \Gamma, Q, q_0, \delta)$ there is a TeSSLa$_\text{bool}$ formula $\phi_R$ and for a TeSSLa$_\text{bool}$ formula $\phi$ there is a DFST $R_\phi = (\Sigma, \Gamma, Q, q_0, \delta)$ s.t.
  \[ \alpha_\Gamma \circ \sem R = \sem{\phi_R} \circ \alpha_\Sigma \quad\text{and}\quad \beta_\Gamma \circ \sem\phi = \sem{R_\phi} \circ \beta_\Sigma.\]
\end{theorem}
Note that since the boolean transducers produce one output symbol per input symbol one could reattach the timestamps of the input streams to the output streams to preserve the exact timestamps, too.

\paragraph{Translating DFST to TeSSLa$_{bool}$}
We represent the states $q \in Q \setminus \{q_0\}$ as stream which is true iff the transducer is in it: $a_q := \mathbf{merge}(x_q,\mathbf{false})$ and the initial state $a_{q_0} := \mathbf{merge}(x_{q_0},\mathbf{true})$, where $x_{q'} := \bigvee_{(a_q,\sigma,a_{q'},\gamma) \in \delta} d_{a_q,\sigma}$.
For every transition $\eta_i = (q,\sigma,q',\gamma)$ we add
$d_{q,\sigma} := \mathbf{last}(a_q, \mathbf{merge}\{s_p \mid p \in \Sigma\}) \land s_\sigma$
and
$o_i := \mathbf{filter}(d_{q,\sigma}, \mathbf{const}(\gamma)(d_{q,\sigma}))$.
The merge of all the output streams is the output:
$\mathit{output} := \mathbf{merge}\{o_i\mid \eta_i \in \delta\}$.

\paragraph{Translating TeSSLa$_\text{bool}$ to DFST}
We translate every equation of the flattened specification $\phi$ into individual DFSTs, which are then composed into one DFST $R_\phi$.
For every DFST the input symbols are functions from the names of the input streams to $\op{Val}$ and
the output symbols are functions from the name of the equation to $\op{Val}$.
As discussed in the previous section, for this finite data domain we only need to consider finitely many different internal states for every equation.
The transition function realizes the state changes the current output based on the current state.
See Appendix~\ref{booltrans} for details on these transducers.

For the composition of the individual DFSTs every two $R = (I \to \op{Val}, O \to \op{Val}, Q, q_0, \delta)$ and $R' = (I' \to \op{Val}, O' \to \op{Val} ,Q' ,q_0' ,\delta')$ are then composed parallel into $R'' = (I \cup I' \to \op{Val},O \cup O' \to \op{Val},Q \times Q',(q_0,q_0'),\delta'')$ with $\delta''((s_1,s_2),g'') = ((s_1',s_2'),h'') \Longleftrightarrow \delta(s_1,g) = (s_1',h) \land \delta'(s_2,g') = (s_2',h') \land g'' = g \cup g' \land \forall \sigma \in I \cap I': g(\sigma) = g'(\sigma) \land h'' = h \cup h'$ until one transducer $R_A = (I_A \to \op{Val},O_A \to \op{Val},Q_A,{q_0}_A,\delta_A)$ represents all equations.
$R_A$ contains transitions with the same in- and output values for certain propositions which represents dependencies between the original equations.
We now build the closure of this transducer which roughly resembles substituting the variables and computing the fixed-point of the equations:
$R_\phi = (I_A \backslash O_A \to \op{Val},O_A \to \op{Val},Q_A,{q_0}_A,\delta_\phi)$,
where $\delta_\phi(s,g) = (s',h) \Longleftrightarrow \delta_A(s,g') = (s',h) \land g = g'|_{I_A \backslash O_A} \land (\forall a \in I_A \cap O_A: g'(a) = h(a))$ for $g|_I := g \cap (I \times \op{Val})$.

Equivalence of deterministic Büchi automata is in P and because the constructed DFSTs can be represented as those we can conclude:
\begin{theorem}
	Equivalence of TeSSLa$_\text{bool}$-formulas is in P.
\end{theorem}

\subsubsection{Timed Fragment}

TeSSLa$_\text{bool+c}$ extends TeSSLa$_\text{bool}$ with the comparison of a timestamp with another, previous timestamp and a constant.
In the syntax, expressions are restricted as follows, where $f \in \mathbb B^n \rightarrowtail \mathbb B$:
\[e := \mathbf{nil} \mid \mathbf{unit} \mid x \mid \mathbf{lift}(f)(e,\dots,e) \mid \mathbf{lift}(g_v)(\mathbf{time}(e), \mathbf{last}(\mathbf{time}(e),e)) \mid \mathbf{last}(e,e)\]
Time comparison is restricted to expressions $\mathbf{lift}(g_v)(\mathbf{time}(a), \mathbf{last}(\mathbf{time}(b),a))$ for streams $a, b \in \mathcal S_{\mathbb B}$ and a constant $v \in \mathbb T$, where $g_v$ is a function $g_v: \mathbb T \times \mathbb T \to \mathbb B$ of the form $g_v(t_1, t_2) = t_1 \lessgtr t_2 + v$ with $\op{\lessgtr} \in \{<, >\}$, which allows checking the temporal distance of the current events of two streams. This is directly related to how clock constraints in timed automata~\cite{timedAutomata,clocks} work.

A \emph{timed finite state transducer (TFST)} is a DFSTs with an additional set of clocks $C$ and $\delta: Q \times \Sigma \times \Theta(C) \rightarrow Q \times 2^C \times \Gamma$ where $\Theta(C)$ is the set of clock constraints. A clock constraint $\vartheta \in \Theta(C)$ is defined over the grammar $\vartheta ::= true \mid T \leq x + c \mid T \geq x + c \mid \neg \vartheta \mid \vartheta \land \vartheta$, where $x \in C$, and $c \in \mathbb T$ is a constant and $T$ refers to the current time.
$\delta$ now also takes a clock constraint and provides a set of clocks that have to be reset to $T$ when taking this transition.
A run of a TFST extends a run of a DFST with timestamps in the input and output word. An additional clock constraint has to be fulfilled to take a transitions and when taking a transitions, some clocks are set to the current time $T$. A more formal definition can be found in Appendix~\ref{timedTransducers}.

TFSTs resemble timed automata using the notion of clock constraints from~\cite{clocks}.
A TFST is called \textit{deterministic}, or DTFST, iff for any two different transitions $\eta_1, \eta_2 \in \delta$ their conjuncted clock constraints $\vartheta_{\eta_1} \land \vartheta_{\eta_2}$ are unsatisfiable.

To show that TeSSLa$_\text{bool+c}$ and DTFSTs have the same expressiveness, we again encode words as streams and vice versa, but this time 
$\alpha_\Sigma$ and $\beta_\Sigma$ preserve the timestamps. Hence both representations are now isomorphic and we can use the inverse encoding functions for decoding:
\begin{theorem}
  For a DTFST $R = (\Sigma, \Gamma, Q, q_0, C, \delta)$ a TeSSLa$_\text{bool+c}$ formula $\phi_R$ exists and for a TeSSLa$_\text{bool+c}$ formula $\phi$ a DTFST $R_\phi = (\Sigma, \Gamma, Q, q_0, C, \delta)$ exists:
  \[ \sem R = \alpha^{-1}_\Gamma \circ \sem{\phi_R} \circ \alpha_\Sigma \quad\text{and}\quad \sem\phi = \beta^{-1}_\Gamma \circ R_\phi \circ \beta_\Sigma.\]
\end{theorem}

\begin{figure}[t]
  \hskip-.035\textwidth
  \begin{minipage}{.395\textwidth}
    \raisebox{8mm}{

\begin{tikzpicture}[font=\tiny]
  \draw[->, blue]
    (.5,.5) -- (.5,3.8) node[right, inner sep=5pt] {data};
  \node[blue, left, rotate=-12, inner sep=2pt] at (.5,1) {none};
  \node[blue, left, rotate=-12, inner sep=2pt] at (.5,2) {bounded};
  \node[blue, left, rotate=-12, inner sep=2pt] at (.5,3) {unbounded};
  \draw[->, red]
    (.5,.5) -- (3.8,.5) node[above, anchor=west, inner sep=5pt, rotate=90] {timestamps};
  \node[red, below, anchor=west, rotate=-12, inner sep=0pt] at (0.8,0.4) {ordering};
  \node[red, below, anchor=west, rotate=-12, inner sep=0pt] at (1.8,0.4) {comparison};
  \node[red, below, anchor=west, rotate=-12, inner sep=0pt] at (2.8,0.4) {creation};
  \filldraw[fill=blue!50!red!50,draw=blue!50!red]
    (.5,.5) rectangle (3.5,3.5);
  \filldraw[fill=blue!75!red!50,draw=blue!75!red]
    (.5,.5) rectangle (2.5,3.5);
  \filldraw[fill=blue!25!red!50,draw=blue!25!red]
    (.5,.5) rectangle (3.5,2.5);
  \filldraw[fill=blue!50!red!25,draw=blue!50!red]
    (.5,.5) rectangle (2.5,2.5);
  \filldraw[fill=blue!25!red!25,draw=blue!25!red]
    (.5,.5) rectangle (2.5,1.5);
  \filldraw[fill=blue!50!red!12,draw=blue!50!red]
    (.5,.5) rectangle (1.5,1.5);
  \node[text width=3cm,align=center] at (1,1) {\textbf{bool\\TeSSLa}\\[3pt]DFST};
  \node[text width=3cm,align=center] at (2,1) {\textbf{bool+c\\TeSSLa}\\[3pt]TFST};
  \node[text width=3cm,align=center, font=\small] at (1.5,2) {TeSSLa};
  \node[text width=3cm,align=center,rotate=90,font=\small] at (3,1.5) {TeSSLa\\+delay};
  \node[text width=3cm,align=center,font=\small] at (1.5,3) {TeSSLa\\+Data};
  \node[text width=3cm,align=center] at (3,3) {\textbf{full\\TeSSLa}};
\end{tikzpicture}}
  \end{minipage}
  \begin{minipage}{.64\textwidth}
    \caption{
      TeSSLa fragments are restricted regarding
      a) event values and available \emph{data} structures and
      b) event \emph{timestamps} and how events sequences are recognized and generated:
      TeSSLa$_\text{bool}$ only checks event \emph{ordering} like deterministic Büchi automata and BSRV~\cite{bozzelli14foundations} (LOLA restricted to boolean streams).
      TeSSLa$_\text{bool+c}$ additionally has timestamp \emph{comparison} with constants like deterministic timed automata.
      TeSSLa has arbitrary \emph{bounded} data structures and arbitrary computations on the timestamps.
      Full TeSSLa allows \emph{unbounded} data structures and the creation of new timestamps via \textbf{delay}.
    }
    \label{figure:tessla-expressiveness-grid}
    \vskip5ex\strut
  \end{minipage}
  \vskip-13ex\strut
\end{figure}
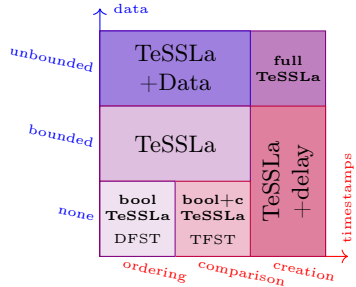

\paragraph{Translating DTFST to TeSSLa$_\text{bool+c}$}
We reuse the translation for DFSTs with the following adjustments: 
We extend the stream $d_{q,\sigma}$ to $d_{q,\sigma,\vartheta}$ by adding the timing constraint $\vartheta$, which is translated by lifting the boolean combination to signal semantics and translating the constraint $T \lessgtr x + c$ to
$\mathbf{time}(\mathbf{merge}\{s_p \mid p \in \Sigma\}) \lessgtr \mathbf{last}(\mathbf{time}(\mathbf{merge}(b_x,\mathbf{unit})),\mathbf{merge}\{s_p \mid p \in \Sigma\}) + c$.
Also for every clock $x \in C$ we add
$b_x := \mathbf{merge}\{ \mathbf{filter}(d_{q,\sigma, \vartheta}, d_{q,\sigma, \vartheta}) \mid (q,\sigma, \vartheta,q',r,\gamma) \in \delta \wedge x \in r \}$.

\paragraph{Translating TeSSLa$_\text{bool+c}$ to DTFST}
The transducers from the equations in $\phi$ are build as before, but instead of translating equations that compare timestamps, we now translate equations of the form $\mathbf{lift}(g_v)(\mathbf{time}(a), \mathbf{last}(\mathbf{time}(b),a))$.
Besides the \textbf{lift} and \textbf{last} operators, it also contains a comparison on timestamps, which is translated using the clocks and clock constraints of the DTFSTs to remember and compare timestamps.
A formalization is given in Appendix~\ref{timedtrans}.
The parallel composition algorithm for DFSTs is extended by conjuncting the timing constraints of the composed transducers.
Afterwards the same closure algorithm is applied.
Equivalence of deterministic timed automata is PSPACE-complete~\cite{timedAutomata} and because the constructed DTFSTs can be represented as those we can conclude:
\begin{theorem}
	Equivalence of TeSSLa$_\text{bool+c}$-formulas is PSPACE-complete.
\end{theorem}

\noindent \autoref{figure:tessla-expressiveness-grid} shows the modularity of the different TeSSLa fragments.


\section{TeSSLa Implementations and Tool Support}
\label{sec:implementation}

The TeSSLa semantics presented in this paper allows multiple implementation styles: Centralized implementations using global memory which take one synchronized input word, as well as distributed implementations using message passing which take individual asynchronous input streams.

Centralized implementations are based on the same idea as the transducers: A global step function triggers the reevaluation of all TeSSLa operators involved in the specification for one timestamp, i.e. until a delayed-labelled edge in the dependency graph is reached. This step function is either triggered by new input events or a timeout of a delay if that has a smaller timestamp. Therefore every delay can register its timeouts globally s.t. the programs main loop can check with every incoming new events if the step function must be triggered for earlier delays before handling the external input. This implementation form is well-suited for software implementations running on traditional CPUs because it minimizes the internal communication overhead. Because software implementations can use dynamic memory management, the integration of unbounded data structures is straightforward. A formalization of such transducers is given in Appendix~\ref{sec:TT}.

As motivated in the introduction, one goal of TeSSLa's design is to allow distributed, parallel implementations with finite memory, e.g. on embedded systems or FPGAs. In this scenario we neither have dynamic memory management nor can we implement a global step function. Instead, every operator in the dependency graph is translated into a computation node with a fixed-size memory cell and finite input queues storing incoming events for every dependency. This setup has already been discussed for a preliminary non-recursive version of TeSSLa in~\cite{oldtessla}. The streams used in the TeSSLa semantics presented in this paper have an explicit notion of progress, which allows the local composition of TeSSLa operators without a global synchronization. Hence every computation node can produce a new output value if at least one input queue contains a new event and all other input queues contain at least progress until the timestamp of this event. The output value is sent to the input queues of all nodes depending on this node. While recursive equations in the transducers are solved by building the closure of the transducer created by applying the parallel composition to all computation nodes, in this message passing scenario we actually implement the Kleene chain of the fixed-point defined in the TeSSLa semantics in \autoref{def:tessla-semantics}: Progress and values are circulated in the cyclic graph of computation nodes until the progress increases no longer, which is exactly when the fixed point is reached. Since every computation node only produces new output events if there is enough progress on every input queue, we can guarantee that the fixed point is computed before new external events are processed.

For practical evaluations of TeSSLa we implemented a TeSSLa compiler in Scala which parses the TeSSLa specification, performs static type checking and converts the specification to flat TeSSLa. Additionally we added a macro system to be able to specify more complex functions based on the basic TeSSLa operators. The macro system allows to build application-domain-specific standard libraries, which makes TeSSLa a very flexible and powerful but still convenient and easy-to-learn specification language.

Furthermore, the types of the input streams are declared explicitly and the user can specify which streams should be contained in the output. Using the macro system, implicit application of $\mathbf{slift}$ to functions and implicit conversion from constants to constant signals, we can write the event counting example from the introduction as follows:

\vskip-2ex
{\centering
\begin{minipage}[t]{8cm}
  \begin{lstlisting}[gobble=4,language=tessla,]
    def count[A](a: Events[A]) := {
      def c: Events[Int] := merge(last(c, a) + 1, 0)
      c }
  \end{lstlisting}
\end{minipage}
\begin{minipage}[t]{2.8cm}
  \begin{lstlisting}[gobble=4,language=tessla,]
    in x: Events[Unit]
    def y := count(x)
    out y
  \end{lstlisting}
\end{minipage}\par}

\noindent We combined the compiler with an interpreter written in Scala, which allows the usage of Java data structures. In order to apply TeSSLa for runtime verification we instrument the LLVM byte code of C programs and analyse this trace online with TeSSLa. This tool chain is available as a Docker container and a web IDE\footnote{\url{http://www.tessla.io}}.


\section{Conclusion}
In this paper we presented the real-time specification language TeSSLa which operates on independent, timed streams and proved that it is suitable for online monitoring.
We characterized the expressiveness of TeSSLa in terms of certain classes of stream-transforming functions.
We also proved the equivalence of a boolean and a timed fragment of TeSSLa to respective classes of transducers and thereby obtained that equivalence for those fragments is in P and PSPACE, resp.
These results facilitate advanced optimizations and static analyses of specifications, e.g. whether such a specification can generate certain outputs.
We presented an implementation based on infinite-state transducers and sketched how TeSSLa is also suitable for parallelized implementations.

\newpage

\appendix
\section{Appendix}

\subsection{Extending Binary Lift to $n$-Ary Lift}\label{def:n-ary-lift}

The binary lift can naturally be extended to an $n$-ary lift by recursively combining two streams into a stream partially applied functions until the final result is obtained:

\begin{definition}[$n$-Ary Lift]
  The $n$-ary \emph{lift} is given as
  \[\sem{\mathbf{lift}(f)(e_1, \ldots, e_n)}_{s_1, \ldots, s_k} = \mathsf{lift_n}(f)(\sem{e_1}_{s_1, \ldots, s_k}, \ldots, \sem{e_n}_{s_1, \ldots, s_k})\]
  where 
  {\small
  \begin{align*}
    \mathsf{lift}_n & \in (𝔻_1 × \cdots × 𝔻_n \rightarrowtail 𝔻') → (𝓢_{𝔻_1}×\cdots×𝓢_{𝔻_n}→𝓢_{𝔻'})\\
    \mathsf{lift}_n(f)(w_1,\dots,w_n) &= \mathsf{liftrec}(f)(w_1,\dots,w_n)\\
    \mathsf{liftrec}(f)(w) &= \mathsf{lift}_1(f_1)(w)\\
    \mathsf{liftrec}(f)(w_1,\dots,w_{n-1},w_n) &= \mathsf{lift}_2(f_n)(\mathsf{liftrec}(f)(w_1,\dots,w_{n-1}), w_n)\\
    f_1(s_1) &= λ(x_2, \dots, x_{\mathsf{arity}(f)}).f(s_1, x_2, \dots, x_n)\\
    f_i(f', x_i) &= λ(x_{i+1}, \dots, x_{\mathsf{arity}(f)}).\begin{cases}
      f'(x_i, \dots, x_n)&\text{if } f'≠⊥\\
      f(⊥,\dots⊥,x_i, \dots,x_n)&\text{if } f'=⊥
    \end{cases}\\
    f_{\mathsf{arity}(f)}(f', e) &= \begin{cases}
      f'(e)&\text{if } f'≠⊥\\
      f(⊥,\dots⊥,e)&\text{if } f'=⊥
    \end{cases}
  \end{align*}}
\end{definition}

\subsection{Proofs for Expressiveness Theorems in Section~\ref{sec:properties}} \label{sec:expressiveness-proofs}

\begin{customTheorem}{3}[Expressiveness of TeSSLa Without Delay]\label{thm:expressiveness-no-delay}
  Every function $f \in \mathcal S_{\mathbb D_1} \times \ldots \times \mathcal S_{\mathbb D_k} \to \mathcal S_{\mathbb D'_1} \times \ldots \times \mathcal S_{\mathbb D'_n}$ on streams can be represented as a TeSSLa specification without delay iff it is
  \begin{inparaenum}[a)]
    \item monotonic and continuous,
    \item timestamp conservative and
    \item future independent.
  \end{inparaenum}
\end{customTheorem}

\begin{proof}
  The function $f$ can be represented as the iterative function $\tilde f \in \mathbb M \cup \{\bot\} \times (\mathbb D_1 \cup \{\bot\} \times \ldots \times \mathbb D_k \cup \{ \bot \})^n \times \mathbb T \to \mathbb M$ with $\tilde f(m, d, t) = m'$, which takes a memory state $m$, the current input values $d$, and the corresponding current timestamp $t$ and returns the new memory state $m'$. Based on this memory state the function $\tilde o \in \mathbb M \to \mathbb D'_1 \cup \{\bot\} \times \ldots \times \mathbb D'_n \cup \{\bot\}$ provides the output events for all output streams.
  For a single input stream $x \in \mathcal S_{\mathbb D}$ and a single output stream $y \in \mathcal S_{\mathbb D}$ the relation between $f$ and its iterative step function $\tilde f$ and the output function $\tilde o$ can be described using an inductively defined function $g \in \mathbb M \cup \{\bot\} \times \mathcal S_{\mathbb D} \to \mathcal S_{\mathbb D}$, s.t. $f(x) = y = g(\bot, x)$. $g$ executes $\tilde f$ and $\tilde o$ for every event of the input stream: 
  $g(m, tdw) = t\tilde o(m')g(m', w)$ and $g(m, t) = t$ where $m' := \tilde f(m, d, t)$.
  $g$ can be naturally extended to tuples of input and output streams by synchronizing the events by their timestamps and fill in missing events with $\bot$.
  Because $f$ is monotonic it is sufficient to compute the output events step by step because $f$ is future independent it is sufficient to allow $\tilde f$ to store arbitrary information about the past events and because $f$ is timestamp conservative it is sufficient to execute $\tilde f$ for every timestamp in the input events.
  Using this representation we can translate $f$ into the following equivalent TeSSLa specification with input variables $x_1, \ldots, x_k$ and output variables $y_1, \ldots, y_n$:
  \begin{align*}
    t &:= \mathbf{time}(\mathbf{merge}(x_1, \ldots, x_k)) \\
    m &:= \mathbf{lift}(\tilde f)(\mathbf{last}(m, t), x_1, \ldots, x_k, t) \\
    \forall_{i \le n}\ y_i &:= \mathbf{lift}(\tilde o_i)(m)
  \end{align*}
  The other direction follows immediately from the fact that every TeSSLa specification without delay is monotonic and continuous, timestamp conservative and future independent.
\end{proof}

\begin{customTheorem}{4}[Expressiveness of TeSSLa With Delay]\label{thm:expressiveness-delay}
  Every function $f \in \mathcal S_{\mathbb D_1} \times \ldots \times \mathcal S_{\mathbb D_k} \to \mathcal S_{\mathbb D'_1} \times \ldots \times \mathcal S_{\mathbb D'_n}$ can be represented as a TeSSLa specification with delay iff it is
  \begin{inparaenum}[a)]
    \item monotonic and continuous and
    \item future independent.
  \end{inparaenum}
\end{customTheorem}

\begin{proof}
  By removing the timestamp conservatism requirement we need to accompany the step-function $\tilde f$ and the output function $\tilde o$ with a timeout function $\tilde u$ which is evaluated on every new memory state and can return the timestamp of the next evaluation of $\tilde f$ if no input event happens before. One timeout function is sufficient because the step-function $\tilde f$ can perform arbitrary computations and store arbitrary state in order to simulate multiple timeouts. The inductively defined function $g \in \mathbb M \cup \{\bot\} \times \mathcal S_{\mathbb D} \to \mathcal S_{\mathbb D}$ executes $\tilde f$ and $\tilde o$ now for every event of the input stream and every due timeout, s.t. $f(x) = y = g(\bot, x)$:
  \begin{align*}
    g(m, tdw) &= \begin{cases}
      t \tilde o(m') g(m', w) \text{ with } m' := \tilde f(m, d, t) &\text{if } t \le \tilde u(m) \\
      \tilde u(m) \tilde o(m') g(m', tdw) \text{ with } m' := \tilde f(m, \bot, \tilde u(m)) &\text{otherwise }
    \end{cases}\\
    g(m, t) &= \begin{cases}
      t &\text{if } t \le \tilde u(m) \\
      \tilde u(m)\tilde o(m')g(m', t) \text{ with } m' := \tilde f(m, \bot, \tilde u(m)) &\text{otherwise}
    \end{cases}
  \end{align*}
  Using this representation we can translate $f$ into the following equivalent TeSSLa specification with input variables $x_1, \ldots, x_k$ and output variables $y_1, \ldots, y_n$:
  \begin{align*}
    t &:= \mathbf{time}(\mathbf{merge}(x_1, \ldots, x_k, \mathbf{delay}(d, \mathbf{merge}(x_1, \ldots, x_k)))) \\
    d &:= \mathbf{time}(t) - \mathbf{lift}(\tilde u)(m) \\
    m &:= \mathbf{lift}(\tilde f)(\mathbf{last}(m, t), x_1, \ldots, x_k, t) \\
    \forall_{i \le n}\ y_i &:= \mathbf{lift}(\tilde o_i)(m) \qedhere
  \end{align*}
\end{proof}

\subsubsection{Constructive Delay Elimination}\label{sec:delay-elimination-proof}
The following lemma and its proof show constructively how to validate the delay operator in a special case of the more general \autoref{thm:expressiveness-no-delay}.

\begin{lemma}[Delay Elimination]
  For every TeSSLa specification $\phi$ with $\sem\phi \in \mathcal S_{\mathbb D_1} \times \ldots \times \mathcal S_{\mathbb D_k} \to \mathcal S_{\mathbb U} \times \mathcal S_{\mathbb D'_1} \times \ldots \times \mathcal S_{\mathbb D'_n}$ with one delay operator there exists a TeSSLa specification $\phi'$ without a delay operator, which derives a boolean stream $z \in \mathcal S_{\mathbb B}$, s.t. for any input streams $x_1,\ldots,x_k$ and any output streams $y_1, \ldots, y_n$ and any stream $d \in \mathcal S_{\mathbb U}$ we have $\sem\phi(x_1, \ldots, x_k) = d, y_1, \ldots, y_n$ iff $\sem{\phi' |_z}(x_1, \ldots, x_k, d, y_1, \ldots, y_n)$ where $d$ is the stream derived immediately through the delay operator.
\end{lemma}

\begin{proof}
  We construct the validating specification $\phi'$ from the generating specification $\phi$:
  We show how to convert the TeSSLa specification $\phi$, which derives new output streams, into the TeSSLa specification $\phi'$, which validates that the output streams were derived according to $\phi$.
  The specification $\phi'$ derives all the streams except $d$ in the same way as $\phi$ does.
  The stream $d$ was derived in $\phi$ using the delay operator and is provided as input to $\phi'$.
  The specification $\phi'$ now validates if the derived streams are equivalent to the input streams $y_1, \ldots, y_n$.
  In order to check two streams for equivalence one has to assert the equivalence of the timestamps and the values separately, i.e. for every two streams $y,y' \in \mathcal S_{\mathbb D}$ we assert $\op{time}(y) = \op{time}(y') \wedge y = y'$, where $=$ denotes the lifted equal function on $\mathbb D$.
  Next we have to assert the correct derivation of $d$:
  Assuming that $\phi$ contains $d = \mathbf{delay}(a,r)$ we derive the effective delays in absolute time by distinguishing three cases:
  We accept the new delay on $a$ either when the old delay is due or together with an event on $r$. Events on $r$ without a new delay on $a$ can reset the effective delay to $\infty$:
  \begin{align*}
    a' := \mathbf{merge}(\ &\mathbf{filter}(\mathbf{last}(a', a) = \mathbf{time}(a), \mathbf{time}(a) + a),\\
    & \mathbf{filter}(\mathbf{time}(r) = \mathbf{time}(a), \mathbf{time}(a) + a),\\
    & \mathbf{const}(\infty)(\mathbf{merge}(r, \mathbf{unit}))\ )
  \end{align*}
  Now we assert at all events $t := \mathbf{merge}(a, r, d)$ that either the delay was fulfilled by an event on $d$, i.e.
  $\mathbf{merge}(\mathbf{time}(d), \mathbf{zero}) = \mathbf{last}(a', t)$
  or the delay is not due and there was no event on $d$, i.e.
  $\mathbf{time}(t) > \mathbf{merge}(\mathbf{time}(d), \mathbf{zero}) \wedge \mathbf{last}(a', t) > \mathbf{time}(t).$
  The boolean stream $z$ is derived as conjunction of all these assertions.
\end{proof}

\subsection{Translating TeSSLa$_\text{bool}$-Operators to DFST} \label{booltrans}

 The function toDFST builds the transducers for a single equation of a TeSSLa$_\text{bool}$-formula $\phi$:\\
 $\mathbf{toDFST}(z := \mathbf{nil}) = (\emptyset,\{z\} \to \op{Val},\{s\},s,\delta)$ with 
 	$\delta(s,\emptyset) = (s,\{z \mapsto \bot\})$\\
 $\mathbf{toDFST}(z := \mathbf{unit}) = (\emptyset,\{z\} \to \op{Val},\{s_0,s_1\},s_0,\delta)$ with\\
 $\delta(s_0,\emptyset) = (s_1,\{z \mapsto \testrue \})$ and $\delta(s_1,\emptyset) = (s_1,\{z \mapsto \bot\})$.\\
 $\mathbf{toDFST}(z := \mathbf{lift}(f)(a^0,\dots,a^n)) = (\{a^{i\le n}\} \to \op{Val},\{z\} \to \op{Val}, \{s,s_e\},s,\delta)$ consists of two states: $s$ where the transducer loops for every input and applies $f$ to the inputs to compute the output and the sink $s_e$ which is reached when one of the input streams ends:\\

 $\delta(s, h) = \begin{cases}
   (s,\{z \mapsto g(b^0, \dots, b^n)\}) & \text{if } \forall i. h(a^i) \in \{\testrue, \tesfalse, \bot\} \\
   (s_e, \{z \mapsto \op{<}'\}) & \text{if } \exists i. h(a_i) = \op{<}'\\
   (s_e,\{z \mapsto g(b^0, \dots, b^n)'\} ) & \text{otherwise},
 \end{cases}$\\
 where
 $b^i = \begin{cases}
   \testrue & \text{if } h(a^i) \in \{\testrue, \testrue' \} \\
   \tesfalse & \text{if } h(a^i) \in \{\tesfalse, \tesfalse' \} \\
   \bot & \text{otherwise}
 \end{cases}$ \\
 and $g(b^0, \dots, b^n) = \begin{cases}
   \bot &\text{if } \forall i. b^i = \bot \\
   f(b^0, \dots, b^n) &\text{otherwise}
 \end{cases}$

 $\mathbf{toDFST}(z := \mathbf{last}(a,b))$ has three states: It stays in the initial state as long as no event on $a$ occurred. The other two states remember the last value that occurred on $a$. For every trigger event on $b$, the value of the current state is the output. Three additional states ensure the proper handling of inclusive and exclusive stream endings.
 
 \noindent $\mathbf{toDFST}(z := \mathbf{last}(a,b)) = (\{a, b\} \to \op{Val}, \{z\} \to \op{Val},\{s_0,s_\testrue,s_\tesfalse,s_w,s_v,s_e\},s_0,\delta)$.
 For the definition $\delta$ we use the following abbreviations: $\testf = \{\testrue, \tesfalse\}$ and $a_x = \{a \mapsto x\}$ for $x \in \{\bot, \testrue, \tesfalse\}$ and $a'_x = \{a \mapsto x'\}$ for $x \in \{<, \bot, \testrue, \tesfalse\}$ and $a^?_x = \{a \mapsto x\}$ for $x \in \op{Val}$:
 \begin{align*}
 	\delta(s_0, a_\bot \cup b_y) &= (s_0, z_\bot)\\
 	\delta(s_0, a_{x \in \testf} \cup b_y) &= (s_x, z_\bot)\\
 	\delta(s_0, a'_x \cup b'_y) &= (s_e, z'_\bot)\\
 	\delta(s_0, a_x \cup b'_y) &= (s_w, z_\bot)\\
 	\delta(s_0, a'_x \cup b_y) &= (s_v, z_\bot)\\
 	\delta(s_{d \in \testf}, a_\bot \cup b_\bot) &= (s_d, z_\bot)\\
 	\delta(s_{d \in \testf}, a_{x \in \testf} \cup b_\bot) &= (s_x, z_\bot)\\
 	\delta(s_{d \in \testf}, a_\bot \cup b_{y \in \testf}) &= (s_d, z_d)\\
 	\delta(s_{d \in \testf}, a_{x \in \testf} \cup b_{y \in \testf}) &= (s_x, z_d)\\
 	\delta(s_{d \in \testf}, a'_x \cup b_y) &= (s_v, z_\bot)\\
 	\delta(s_{d \in \testf}, a^?_x \cup b'_{y \in \{<,\bot\}}) &= (s_e, z'_y)\\
 	\delta(s_{d \in \testf}, a^?_x \cup b'_{y \in \testf}) &= (s_e, z'_d)\\
 	\delta(s_w, a_\bot \cup b^?_y) &= (s_w, z_\bot)\\
 	\delta(s_w, a_{x \in \testf} \cup b^?_y) &= (s_e, z'_\bot)\\
 	\delta(s_w, a'_x \cup b^?_y) &= (s_e, z'_\bot)\\
 	\delta(s_v, a^?_x \cup b_\bot) &= (s_v, z_\bot)\\
 	\delta(s_v, a^?_x \cup b'_\bot) &= (s_e, z'_\bot)\\
 	\delta(s_v, a^?_x \cup b^?_{y \in \{<, \testrue, \tesfalse\}}) &= (s_e, z'_<)
 \end{align*}

 $\mathbf{toDFST}(z := \mathbf{slift}(\geq)(\mathbf{time}(a), \mathbf{time}(b))$ consists of three initialization states remembering on which input streams an event already occurred. When on both streams an event occurred the transducer transits in a fourth state in which it stays until one stream ends. In this loop the output is true for every event on $a$ and false for events occurring only on $b$:

 \noindent $\mathbf{toDFST}(z := \mathbf{slift}(\geq)(\mathbf{time}(a), \mathbf{time}(b))) =$\\
 \strut\qquad $ (\{a, b\} \to \op{Val},\{z\} \to \op{Val},\{s_\bot,s_a,s_b,s_c,s_e\},s_\bot,\delta)$ with
 \begin{align*}
 	\forall s: \delta(s,a_\bot \cup b_\bot) &= (s,z_\bot)\\
 	\forall s: \delta(s,a'_< \cup b^?_{y \not= <}) &= (s_e,z'_<)\\
 	\forall s: \delta(s,a^?_{x \not= <} \cup b'_<) &= (s_e,z'_<)\\
 	\forall s: \delta(s,a'_\testf \cup b_\testf) &= \delta(s,a_\testf \cup b'_\testf) = (s_e,z'_\testrue)\\
 	\forall s: \delta(s,a'_\testf \cup b'_\testf) &= (s_e,z'_\testrue)\\[1ex]
 	\delta(s_\bot,a_{x \in \testf} \cup b_\bot) &= (s_a,z_\bot)\\
 	\delta(s_\bot,a_\bot \cup b_{y \in \testf}) &= (s_b,z_\bot)\\
 	\delta(s_\bot,a_\testf \cup b_\testf) &= (s_c,z_\testrue)\\
 	\delta(s_\bot,a'_{x \neq <} \cup b_\bot) &= \delta(s_\bot,a'_\bot \cup b_\testf) = (s_e,z'_\bot)\\
 	\delta(s_\bot,a'_\bot \cup b'_\testf) &= (s_e,z'_\bot)\\
 	\delta(s_\bot,a_\bot \cup b'_{y \not= <}) &= \delta(s_\bot,a_\testf \cup b'_\bot) = (s_e,z'_\bot)\\
 	\delta(s_\bot,a'_\testf \cup b'_\bot) &= (s_e,z'_\bot)\\[1ex]
 	\delta(s_a,a_\testf \cup b_\bot) &= (s_a,z_\bot)\\
 	\delta(s_a,a_\bot \cup b_\testf) &= (s_c,z_\tesfalse)\\
 	\delta(s_a,a_\testf \cup b_\testf) &= (s_c,z_\testrue)\\
 	\delta(s_a,a'_{x \not= <} \cup b_\bot) &= (s_e,z'_\bot)\\
 	\delta(s_a,a'_\bot \cup b_\testf) &= \delta(s_a,a'_\bot \cup b'_\testf) = (s_e,z'_\tesfalse)\\
 	\delta(s_a,a_\bot \cup b'_\testf) &= (s_e,z'_\tesfalse)\\
 	\delta(s_a,a^?_{x \neq <} \cup b'_\bot) &= (s_e,z'_\bot)\\[1ex]
 	\delta(s_b,a_\bot \cup b_\testf) &= (s_b,z_\bot)\\
 	\delta(s_b,a_\testf \cup b^?_y) &= (s_c,z_\testrue)\\
 	\delta(s_b,a_\bot \cup b'_{y \not= <}) &= (s_e,z'_\bot)\\
 	\delta(s_b,a'_\bot \cup b^?_{y \neq <}) &= (s_e,z'_\bot)\\
 	\delta(s_b,a'_\testf \cup b_\bot) &= (s_e,z'_\testrue)\\
 	\delta(s_b,a_\testf \cup b'_\bot) &= \delta(s_b,a'_\testf \cup b'_\bot) = (s_e,z'_\testrue)\\[1ex]
 	\delta(s_c,a_\bot \cup b_\testf) &= (s_c,z_\tesfalse)\\
 	\delta(s_c,a_\testf \cup b^?_y) &= (s_c,z_\testrue) \\
 	\delta(s_c,a'_\bot \cup b_\bot) &= \delta(s_c,a_\bot \cup b'_\bot) = (s_e,z'_\bot)\\
 	\delta(s_c,a'_\bot \cup b_\testf) &= \delta(s_c,a'_\bot \cup b'_\testf) = (s_e,z'_\tesfalse)\\
 	\delta(s_c,a'_\testf \cup b_\bot) &= (s_e,z'_\testrue)\\
 	\delta(s_c,a_\bot \cup b'_\testf) &= (s_e,z'_\tesfalse)\\
 	\delta(s_c,a_\testf \cup b'_\bot) &= \delta(s_c,a'_\testf \cup b'_\bot) = (s_e,z'_\testrue)
 \end{align*}

\subsection{Timed Transducers}\label{timedTransducers}

A \emph{timed finite state transducer (TFST)} is a 6-tuple $R = (\Sigma, \Gamma, Q, q_0, C, \delta)$ which extends DFSTs with a set of clocks $C$ and $\delta: Q \times \Sigma \times \Theta(C) \rightarrow Q \times 2^C \times \Gamma$ where $\Theta(C)$ is the set of clock constraints. A clock constraint $\vartheta \in \Theta(C)$ is defined over the grammar $\vartheta ::= true \mid T \leq x + c \mid T \geq x + c \mid \neg \vartheta \mid \vartheta \land \vartheta$, where $x \in C$, and $c \in \mathbb T$ is a constant and $T$ refers to the current time.
$\delta$ now also takes a clock constraint and provides a set of clocks that have to be reset to $T$ when taking this transition.
For an input word $w = (t_0,w_0)(t_1,w_1)(t_2,w_2)\dots$ we call a sequence
$s_0,v_0 \xrightarrow[o_0]{(t_0,w_0),\vartheta_0,r_0} s_1,v_1 \xrightarrow[o_1]{(t_1,w_1),\vartheta_1,r_1} s_2,v_2 \xrightarrow[o_2]{(t_2,w_2),\vartheta_2,r_2} \cdots$
a run of a TFST $R$ where $v_i: C \rightarrow \mathbb{R}$ are functions mapping every clock to its current value iff $s_0 = q_0$, $\forall c \in C: v_0(c) = 0$ and for all $i \geq 0$, $\delta(s_i,w_i,\vartheta_i) = (s_{i+1},r_i,o_i)$, $v_{i+1} = v_i[c \leftarrow t_i]$ for all $c \in R$ and $t_i,v_i \models \vartheta_i$ (which means $\vartheta_i$ resolved to \emph{true} if $T$ is replaced with $t_i$ and the corresponding values from $v_i$ are used for the clocks). The output of such a run is the sequence $\sem R(w) = t_0o_0t_1o_1t_2o_2\ldots \in (\mathbb T \times \Gamma)^\infty$.
TFSTs resemble timed automata using the notion of clock constraints from~\cite{clocks}.
A TFST is called \textit{deterministic}, or DTFST, iff for any two different transitions $\eta_1, \eta_2 \in \delta$ their conjuncted clock constraints $\vartheta_{\eta_1} \land \vartheta_{\eta_2}$ are unsatisfiable.
 
\subsection{Translating TeSSLa$_\text{bool+c}$ to DTFST} \label{timedtrans}
 
 The transducers translated from a single equation in a TeSSLa$_\text{bool+c}$-formula $\phi$ are build as before. Instead of the timestamp comparison we translate the new type of equation
 $\op{toDTFST}(z := \mathbf{lift}(g_v)(\mathbf{time}(e), \mathbf{last}(\mathbf{time}(a),e))) = {}$\\
 $\quad (\{a,e\} \to \op{Val}, \{z\} \to \op{Val},\{s_0,s_v,s_e\},s_0,\{c_a\},\delta)$ with $g_v(t_1, t_2) = t_1 \lessgtr t_2 + v$:
 \begin{align*}
   \delta(s_0,e_\bot \cup a_\bot, \op{true}) &= (s_0,\emptyset,z_\bot)\\
   \delta(s_0,e_\bot \cup a_{y\in\testf}, \op{true}) &= (s_0,\{c_a\},z_\bot)\\
   \delta(s_0,e_\bot \cup a'_y,\op{true}) &= (s_v,\emptyset,z_\bot)\\
   \delta(s_0,e_{x\in\testf} \cup a_\bot, T\lessgtr c_a + v) &= (s_0,\emptyset,z_\testrue)\\
   \delta(s_0,e_{x\in\testf} \cup a_\bot, \neg( T\lessgtr c_a + v)) &= (s_0,\emptyset,z_\tesfalse)\\
   \delta(s_0,e_{x\in\testf} \cup a_{y\in\testf}, T\lessgtr c_a + v) &= (s_0,\{c_a\},z_\testrue)\\
   \delta(s_0,e'_{x \in \{<,\bot\}} \cup a^?_y, \op{true}) &= (s_e,\emptyset,z'_x)\\
   \delta(s_0,e'_{x \in \testf} \cup a^?_y, T\lessgtr c_a + v) &= (s_e,\emptyset,z'_\testrue)\\
   \delta(s_0,e'_{x \in \testf} \cup a^?_y, \neg( T\lessgtr c_a + v)) &= (s_e,\emptyset,z'_\tesfalse)\\
   \delta(s_v, e_{y \in \testf} \cup a^?_x, \op{true}) &= (s_e, \emptyset, z'_<)\\
   \delta(s_v, e_\bot \cup a^?_x, \op{true}) &= (s_v, \emptyset, z_\bot)\\
   \delta(s_v, e'_\bot \cup a^?_x, \op{true}) &= (s_e, \emptyset, z'_\bot)\\
   \delta(s_v, e'_{y \in \{<, \testrue, \tesfalse\}} \cup a^?_x, \op{true}) &= (s_e, \emptyset, z'_<)
 \end{align*}
 
 This transducer stays in its initial state $s_0$ until one stream ends: For every $a$ it resets the clock to save the value of $a$. For every $b$ it produces an output on $z$ according to the fulfilment of the current clock constraint. The states $s_v$ and $s_e$ handle the stream endings similarly to the transducer for the \textbf{last} operator.

\subsection{TeSSLa Transducer (TT)}\label{sec:TT}

 Finally we introduce TeSSLa transducers as a new transducer type with the same expressiveness as full TeSSLa with arbitrary time computations, an arbitrary data domain over unbounded data structures and the delay operator, which can create new timestamps. 

 To resemble the delay's behaviour we define: A \emph{TeSSLa Transducer (TT)} is a TFST with an extended transition function, which adds $\epsilon$-transitions with clock constrains, which do not consume an input symbol, but still produce an output symbol. These have to be taken when possible but can only be taken once for every point in time. The run of a TT is extended appropriately.

 \begin{definition}[Transition Function of a TT]
 	The transition function of a TT is $\delta: Q \times (\Sigma \cup \{\epsilon\}) \times \Theta(C) \rightarrow Q \times 2^C \times \Gamma$.
 \end{definition}

 \begin{definition}[Run of a TT]
 	For an input word $w = (t_0,w_0)(t_1,w_1)(t_2,w_2)\ldots$ we call a sequence
 	\[ s_0,v_0\xrightarrow[o_0]{\sigma_0,\tau_0,\vartheta_0,r_0} s_1,v_1 \xrightarrow[o_1]{\sigma_1,\tau_1,\vartheta_1,r_1} s_2,v_2 \xrightarrow[o_2]{\sigma_2,\tau_2,\vartheta_2,r_2} \cdots \]
 	a run of a TT if for all $i \geq 0$: $\sigma_i = \epsilon \land v_i, \tau_i \models \vartheta_i \ \lor\  \sigma_i = w_j \land \tau_i = t_j$ and $\tau_i < \tau_{i+1}$ and $\forall t_j: \exists \tau_k: t_j = \tau_k$ and $\forall t \in (\tau_i, \tau_{i+1}]: t, v_i \models \vartheta_i \Rightarrow t = \tau_{i+1}$, where $v_i: C \rightarrow \mathbb{R}$ are functions mapping every clock to its current value iff $s_0 = q_0$, $\forall c \in C: v_0(c) = 0$ and for all $i \geq 0$, $\delta(s_i,\sigma_i,\vartheta_i) = (s_{i+1},r_i,o_i)$, $v_{i+1} = v_i[c \leftarrow \tau_i]$ for all $c \in r_i$ and $\tau_i,v_i \models \vartheta_i$, i.e. $\vartheta_i$ resolves to true if $T$ is replaced with $\tau_i$ and every $x \in C$ is replaced with the clock's current value $v(x)$. The output of such a run is the sequence $\sem R(w) = \tau_0o_0\tau_1o_1\tau_2o_2\dots \in (\mathbb{T} \times \Gamma)^\infty$.
 \end{definition}

 We call a TT deterministic (or DTT for short), iff its clock constraints fulfil the same property as the ones of a deterministic TFST.

 \paragraph{Translating TeSSLa to DTT} We extend the value domain in the previous construction to arbitrary data types $\mathbb D$ and hence get $\op{Val}_{\mathbb D} = \{\bot, d, <', \bot', d' \mid d \in \mathbb D\}$.
 In case of the last operator instead of the two states $s_{x \in \testf}$ we now have (possibly infinitely many) states $s_{x \in \mathbb D}$. We no longer need translations for special expressions containing the $\mathbf{time}$ operator, but we can now directly translate the time operator itself and we add the translation of the additional $\mathbf{delay}$ operator. In the following we assume $T$ always to be the corresponding $\tau_i$ shown in the run before:
 $\op{toDTT}(z := \mathbf{time}(a)) = (\{a\} \to \op{Val}_{\mathbb D},\{z\} \to \op{Val}_{\mathbb D},\{s_0,s_e\},s_0,\emptyset,\delta)$:
 \begin{align*}
 	\delta(s_0,a_\bot, \op{true}) &= (s_0,\emptyset, \emptyset)&
 	\delta(s_0,a_{d \neq \bot}, \op{true}) &= (s_0, z_T, \emptyset)\\
 	\delta(s_0,a'_{d \not\in \{<,\bot\} },\op{true}) &= (s_e, z'_T, \emptyset)&
 	\delta(s_0,a'_{d \in \{<,\bot\} },\op{true}) &= (s_e,z'_d, \emptyset)
 \end{align*}

 The initial state of $\op{toDTT}(z = \mathbf{delay}(a,b))$ represents that no delay is set. When a new delay is accepted the transducers transits into the corresponding state storing the delay and resets the clock. An outgoing transition guarded with a clock constrained checks if the delay timed out and produces a unit symbol on $z$. Either a new delay is set simultaneously or the transition goes back to the initial state.
 
 Formally, the transducer is given by $\op{toDTT}(z = \mathbf{delay}(a,b)) = ((\{a\} \to \op{Val}_{\mathbb T}) \cup (\{b\} \to \op{Val}_{\mathbb D}),\{z\}\to\op{Val}_{\mathbb U},\{s,s_e,s^b\} \cup \{s_d, s_d^a, s_d^b \mid d \in \mathbb{T}\},s,\{c_b\},\delta)$ with
 \begin{align*}
 	\delta(s, a_\bot \cup b_\bot, \op{true}) &= (s, z_\bot,\emptyset)\\
 	\delta(s, a_{x\neq\bot} \cup b_{y\neq\bot}, \op{true}) &= (s_x, z_\bot,\{c_b\}\\
 	\delta(s, a_{x\neq\bot} \cup b_\bot, \op{true}) &= (s, z_\bot,\emptyset)\\
 	\delta(s, a_\bot \cup b_{y\neq\bot}, \op{true}) &= (s, z_\bot,\emptyset\\
 	\delta(s, a'_{x \in \mathbb{T}} \cup b_{y\neq\bot}, \op{true}) &= (s^a_x, z_\bot,\{c_b\})\\
 	\delta(s, a'_{x \not\in \mathbb{T}} \cup b_{y\neq\bot}, \op{true}) &= (s^e, z'_\bot,\emptyset)\\
 	\delta(s, a'_x \cup b_\bot, \op{true}) &= (s^e, z'_\bot, \emptyset)\\
 	\delta(s, a_\bot \cup b'_y, \op{true}) &= (s^b, z_\bot, \emptyset)\\
 	\delta(s, a_{x \in \mathbb{T}} \cup b'_\bot, \op{true}) &= (s^b, z_\bot, \emptyset)\\
 	\delta(s, a_{x \in \mathbb{T}} \cup b'_{y \neq \bot}, \op{true}) &= (s^b_x, z_\bot, \{c_b\})\\      
 	\delta(s, a'_x \cup b'_y, \op{true}) &= (s^e, z'_\bot, \emptyset)\\[1ex]
 	\delta(s_d, \epsilon, T = d + c_b) &= (s, z_\unitsym, \emptyset)\\
 	\delta(s_d, \epsilon, T \neq d + c_b) &= (s_d, z_\bot, \emptyset)\\
 	\delta(s_d, a_\bot \cup b_\bot, T = d + c_b) &= (s, z_\unitsym, \emptyset)\\
 	\delta(s_d, a_\bot \cup b_\bot, T \neq d + c_b) &= (s_d, z_\bot, \emptyset)\\
 	\delta(s_d, a_{x\neq\bot} \cup b_\bot, T = d + c_b) &= (s_x, z_\unitsym, \{c_b\})\\
 	\delta(s_d, a_{x\neq\bot} \cup b_\bot, T \neq d + c_b) &= (s_d, z_\bot, \emptyset)\\
 	\delta(s_d, a_\bot \cup b_{x\neq\bot}, T = d + c_b) &= (s, z_\unitsym, \emptyset)\\
 	\delta(s_d, a_\bot \cup b_{x\neq\bot}, T \neq d + c_b) &= (s, z_\bot, \emptyset)\\
 	\delta(s_d, a_{x\neq\bot} \cup b_{y\neq\bot}, T = d + c_b) &= (s_x, z_\unitsym, \{c_b\})\\
 	\delta(s_d, a_{x\neq\bot} \cup b_{y\neq\bot}, T \neq d + c_b) &= (s_x, z_\bot, \{c_b\})\\
 	\delta(s_d, a'_{x \in \mathbb{T}} \cup b_\bot, T = d + c_b) &= (s^a_x, z_\unitsym, \{c_b\})\\
 	\delta(s_d, a'_{x \in \mathbb{T}} \cup b_\bot, T \neq d + c_b) &= (s^a_d, z_\bot, \emptyset)\\
 	\delta(s_d, a'_{x \notin \mathbb{T}} \cup b_\bot, T = d + c_b) &= (s^e, z'_\unitsym, \emptyset)\\
 	\delta(s_d, a'_{x \notin \mathbb{T}} \cup b_\bot, T \neq d + c_b) &= (s^a_d, z_\bot, \emptyset)\\
 	\delta(s_d, a_\bot \cup b'_y, T = d + c_b) &= (s^b, z_\unitsym, \emptyset)\\
 	\delta(s_d, a_\bot \cup b'_{y \in \{<, \bot\}}, T \neq d + c_b) &= (s^b_d, z_\bot, \emptyset)\\
 	\delta(s_d, a_\bot \cup b'_{y \not \in \{<, \bot\}}, T \neq d + c_b) &= (s^b, z_\bot, \emptyset)\\
 	\delta(s_d, a_{x \in \mathbb{T}} \cup b'_y, T = d + c_b) &= (s^b_x, z_\unitsym, \{c_b\})\\
 	\delta(s_d, a_{x \in \mathbb{T}} \cup b'_{y \in \{<, \bot\}}, T \neq d + c_b) &= (s^b_d, z_\bot, \emptyset)\\
 	\delta(s_d, a_{x \in \mathbb{T}} \cup b'_{y \not\in \{<, \bot\}}, T \neq d + c_b) &= (s^b_x, z_\bot, \{c_b\})\\
 	\delta(s_d, a'_x \cup b'_y, T = d + c_b) &= (s^e, z'_\unitsym, \emptyset)\\
 	\delta(s_d, a'_x \cup b'_y, T \neq d + c_b) &= (s^e, z'_\bot, \emptyset)\\[1ex]
 	\delta(s^a_d, \epsilon, T = d + c_b) &= (s^e, z'_\unitsym, \emptyset)\\
 	\delta(s^a_d, \epsilon, T \neq d + c_b) &= (s^a_d, z_\bot, \emptyset)\\
 	\delta(s^a_d, a^?_x \cup b_\bot, T = d + c_b) &= (s^e, z'_\unitsym, \emptyset)\\
 	\delta(s^a_d, a^?_x \cup b_\bot, T \neq d + c_b) &= (s^a_d, z_\bot, \emptyset)\\
 	\delta(s^a_d, a^?_x \cup b_{y\neq\bot}, T = d + c_b) &= (s^e, z'_\unitsym, \emptyset)\\
 	\delta(s^a_d, a^?_x \cup b_{y\neq\bot}, T \neq d + c_b) &= (s^e, z'_\bot, \emptyset)\\   
 	\delta(s^a_d, a^?_x \cup b'_y, T = d + c_b) &= (s^e, z'_\unitsym, \emptyset)\\
 	\delta(s^a_d, a^?_x \cup b'_y, T \neq d + c_b) &= (s^e, z'_\bot, \emptyset)\\[1ex]
 	\delta(s^b, a_\bot \cup b^?_y, \op{true}) &= (s^b, z_\bot, \emptyset)\\
 	\delta(s^b, a_{x \in \mathbb{T}} \cup b^?_y, \op{true}) &= (s^b_x, z_\bot, \emptyset)\\[1ex]
 	\delta(s^b_d, \epsilon, T = d + c_b) &= (s^e, z'_<, \emptyset)\\
 	\delta(s^b_d, \epsilon, T \neq d + c_b) &= (s^b_d, z_\bot, \emptyset)\\
 	\delta(s^b_d, a^?_x \cup b^?_y, T = d + c_b) &= (s^e, z'_<, \emptyset)\\
 	\delta(s^b_d, a_\bot \cup b^?_y, T \neq d + c_b) &= (s^b_d, z_\bot, \emptyset)\\
 	\delta(s^b_d, a'_{x \not\in \mathbb{T}} \cup b^?_y, T \neq d + c_b) &= (s^b_d, z_\bot, \emptyset)\\
 	\delta(s^b_d, a^?_{x \in \mathbb{T}} \cup b^?_y, T \neq d + c_b) &= \begin{cases}
 		(s^b_d, z_\bot, \emptyset) & \text{if } v(c_b) + d \leq T + x\\
 		(s^b_x, z_\bot, \{c_b\}) & \text{otherwise}
 	\end{cases}
 \end{align*}

\end{document}